\documentclass[11pt,a4paper]{article}

\usepackage{jcappub}
\usepackage{multirow}
\pdfoutput=1
\usepackage[utf8]{inputenc}
\usepackage{float}
\usepackage{hyperref}
\usepackage{amsmath}
\usepackage{graphicx}
\usepackage{dcolumn}
\usepackage{bm}
\usepackage{epsfig}
\usepackage{amssymb,latexsym,mathrsfs}
\usepackage{graphicx}
\usepackage{color}

\usepackage{comment}

\definecolor{myred}{RGB}{180,50,28}
\definecolor{myblue}{RGB}{2,50,180}
\definecolor{mygreen}{RGB}{2,150,80}

\hypersetup{colorlinks=true,
            citecolor=blue,
            linkcolor=blue,
            urlcolor=blue}

\newcommand{\be}{\begin{equation}}
\newcommand{\ee}{\end{equation}}
\newcommand{\bea}{\begin{eqnarray}}
\newcommand{\eea}{\end{eqnarray}}

\newcommand{\mpl}{M_{*}^{2}}
\newcommand{\bra}{\alpha_{\textrm{B}}}
\newcommand{\run}{\alpha_{\textrm{M}}}
\newcommand{\kin}{\alpha_{\textrm{K}}}
\newcommand{\ten}{\alpha_{\textrm{T}}}

\newcommand{\csN}{c_{\mathrm{sN}}^{2}}

\newcommand{\dd}{{\rm{d}}}
\newcommand{\ii}{{\rm{i}}}

\newcommand{\lcdm}{$\Lambda$CDM}

\newcommand{\CLASS}{\textsc{class}}
\newcommand{\hiclass}{{\tt hi\_class}}
\newcommand{\deltanb}{\delta^{\text{Nb}}}

\newcommand{\deltanbprime}{\delta^{\prime\text{Nb}}}
\newcommand{\deltanbprimeprime}{\delta^{\prime\prime\text{Nb}}}
\newcommand{\deltaN}{\delta^{\text{N}}}

\newcommand{\deltaNprime}{\delta^{\prime\text{N}}}
\newcommand{\deltaNprimeprime}{\delta^{\prime\prime\text{N}}}
\newcommand{\gammanb}{\gamma^{\text{Nb}}}
\newcommand{\HTnb}{H_{\text{T}}^{\text{Nb}}}

\newcommand{\HTnbprime}{H_{\text{T}}^{\prime\text{Nb}}}
\newcommand{\HTnbprimeprime}{H_{\text{T}}^{\prime\prime\text{Nb}}}

\newcommand{\drhoGR}{\delta \rho_{\mathrm{GR}}}

\title{Fully relativistic predictions in Horndeski gravity from standard Newtonian N-body simulations}
\author[a,b,1]{Guilherme Brando,\note{Corresponding author.}}
\author[b]{Kazuya Koyama,}
\author[b]{David Wands,}
\author[c]{Miguel Zumalac\'arregui,}
\author[d]{Ignacy Sawicki,}
\author[e,f]{and Emilio Bellini}

\affiliation[a]{PPGCosmo, CCE -- Universidade Federal do Esp\'irito Santo,\\
Avenida Fernando Ferrari
514, 29075-910 Vit\'oria, Esp\'irito Santo, Brazil}
\affiliation[b]{Institute of Cosmology and Gravitation, University of Portsmouth\\ Dennis Sciama Building, Burnaby Road, Portsmouth PO1 3FX, United Kingdom}
\affiliation[c]{Max Planck Institute for Gravitational Physics (Albert Einstein Institute) \\
Am Mühlenberg 1, D-14476 Potsdam-Golm, Germany}
\affiliation[d]{CEICO, FZU - Institute of Physics of the Czech Academy of Sciences, \\
Na Slovance 1999/2, 182 21 Praha, Czechia
}
\affiliation[e]{Département de Physique Théorique, Université de Genève,\\ 24 quai Ernest Ansermet, 1211 Genève4, Switzerland
}
\affiliation[f]{Oxford Astrophysics, Department of Physics, Keble Road,\\
Oxford, OX1 3RH, U.K
}

\emailAdd{gbrando@cosmo-ufes.org}
\emailAdd{kazuya.koyama@port.ac.uk}
\emailAdd{david.wands@port.ac.uk}
\emailAdd{miguel.zumalacarregui@aei.mpg.de}
\emailAdd{ignacy.sawicki@outlook.com}
\emailAdd{emilio.bellini@unige.ch}

\abstract{The N-body gauge allows the introduction of relativistic effects in Newtonian cosmological simulations. Here we extend this framework to general Horndeski gravity theories, and investigate the relativistic effects that the scalar field introduces in the matter power spectrum at intermediate and large scales. In particular, we show that the kineticity function at these scales enhances the amplitude of the signal of contributions coming from the extra degree of freedom. Using the Quasi-Static Approximation (QSA), we separate modified gravity effects into two parts: one that only affects small-scale physics, and one that is due to relativistic effects. This allows our formalism to be readily implemented in modified gravity N-body codes in a straightforward manner, e.g.,
relativistic effects can be included as an additional linear density field in simulations. We identify the emergence of gravity acoustic oscillations (GAOs) in the matter power spectrum at large scales, $k \sim 10^{-3}-10^{-2}$ Mpc$^{-1}$. GAO features have a purely relativistic origin, coming from the dynamical nature of the scalar field. GAOs may be enhanced to detectable levels by the rapid evolution of the dark energy sound horizon in certain modified gravity models and can be seen as a new test of gravity at scales probed by future galaxy and intensity-mapping surveys.}
\begin{document}
\maketitle
\flushbottom

\section{Introduction}

The current standard model of cosmology, \lcdm, assumes General Relativity (GR) as the description of gravity on all scales. While Einstein's theory continues to pass many local and astrophysical tests, the lack of precision in cosmological tests of gravity when compared to local ones still leaves room for further exploration of modified gravity models. These theories are able to describe cosmological observations and in particular can offer an alternative explanation for the late-time acceleration of our Universe. Even though we end up enlarging the parameter space of the gravitational sector when compared to GR, research into these models occupies a central role in beyond-\lcdm \ models.

Two stage-IV large-scale structure (LSS) surveys will release their first data (DESI~\cite{desi}) or come online (Euclid~\cite{euclid}) in the next few years. Over the last decade much research in cosmology has been devoted to preparing for this moment. It has focused on a variety of topics, such as forecasts for future constraints on cosmological parameters~\cite{euclid_forec}, perfecting analysis pipelines for the huge amount of data, and generation of state-of-the-art simulation suites~\cite{bacco,outerrim,aemulus,abacus,mice,coyote,Uchuu}. While in most cases the theoretical background for this work was the standard cosmological model, and minimal extensions to it, modified gravity has also seen considerable advances, especially in Newtonian N-body simulation codes~\cite{farbodlombr,mg_nb,mg_cola}. This has pushed the field to the same precision requirements as the investigation of \lcdm \ cosmology, and further developing simulation techniques in modified gravity is still of great importance. 

Most N-body simulations, however, do not capture the relativistic nature of our Universe at large distances, since they are inherently Newtonian. While Newton's gravity is an accurate description of our cosmology at small scales, at large scales relativistic effects come into play, which may lead to deviations from the Newtonian description. Since the matter content of N-body simulations is restricted to cold dark matter species (a pressureless, collisionless component that contributes the majority of the current matter content in our Universe), other relativistic species in the Universe are not captured by these simulations.

To address this problem, one approach is to use relativistic simulations, such as the \emph{gevolution} code~\cite{gevolution}, based on a weak-field expansion in the Poisson gauge. Recently, the same code was generalized to clustering dark energy cosmologies~\cite{kevolution1,kevolution2}, named \emph{k-evolution}. An alternative approach is to construct a relativistic gauge in which Newtonian simulations can be interpreted as a consistent solution of GR, embedded in a perturbed spacetime~\cite{green,chisari,flender,haugg,bruni}. One such gauge is the N-body gauge~\cite{Nbody1,Nbody2,Nbody3,Nbody4,Nbody5,Dakin:2019vnj}, in which the physical number density of particles matches the coordinate number density of particles at first order in Newtonian simulations, i.e., the relativistic particle density does not suffer any deformation in its volume element. Since its conception in~\cite{Nbody1}, a number of other physically relevant gauges have been discussed in the literature, such as the class of Newtonian motion gauges~\cite{Nbody2,Nbody3,Nbody4}, including the N-boisson gauge~\cite{Nboisson}, which combines the spatial threading condition of the N-body gauge with the temporal slicing of the Poisson gauge.
Since the linear theory is valid at large scales, and Newtonian gravity accurately describes our Universe at small scales, the combination of the N-body gauge with Newtonian simulations makes this approach attractive, as it combines the fast and computationally low-cost Einstein-Boltzmann solvers with state-of-the-art Newtonian N-body simulation codes. 

In the present paper we will generalize previous work~\cite{us}, in which we showed how modified gravity effects can be introduced in the N-body gauge, using an effective fluid description. In this work, we will demonstrate and discuss how the full space of scalar-tensor theories described by Horndeski theory~\cite{horn,defa1,koba}, can be implemented in the publicly available Einstein-Boltzmann solver \hiclass~\cite{hiclass1,hiclass2}, the modified gravity version of the General Relativity solver \CLASS~\cite{class1,class2}. We will also analyze the effect and impact of the scalar field in the relativistic matter power spectrum, comparing it to its linear Newtonian counterpart. This analysis is different from previous work investigating modified gravity at very large scales, such as~\cite{baker,kk2,renk,alonso}, which investigated the impact of the scalar field on line-of-sight corrections to galaxy number counts. Our framework can be used to construct these number counts, from the relativistic output of N-body simulations, using ray-tracing techniques.

This paper is structured as follows: in Section~\ref{sec:sec2} we will review the N-body gauge formalism and Horndeski's theory of gravity. Section~\ref{sec:sec3} is devoted to the study of the relativistic effects that the dark energy scalar field introduces in the matter power spectrum, and how we can separate these corrections into purely relativistic ones and contributions that can be described by Newtonian gravity, which are already present in modified gravity N-body codes. We will also analyse non-Newtonian oscillatory features that emerge due to the presence of the scalar field. In Section~\ref{sec:concl} we conclude with our final considerations on the results presented in the previous section.

\section{The N-body gauge and Modified Gravity}\label{sec:sec2}

\subsection{N-body gauge}\label{sec:Nb}
We begin by outlining the mathematical framework of our implementation. We consider scalar perturbations on top of a homogeneous and isotropic Friedmann-Lemaitre-Robertson-Walker (FLRW) background, with the line element given by:
\begin{subequations}\label{eq:metric-potentials}
\begin{align}
  g_{00} &= -a^2 \left( 1 + 2A \right) \,, \\
  g_{0i} &=  a^2\, \ii \hat k_i B \,, \\
  g_{ij} &= a^2 \left[ \delta_{ij} \left( 1 + 2 H_{\mathrm{L}} \right) + 2 \left( \delta_{ij}/3 - \hat{k}_i \hat{k}_j \right) H_{\mathrm{T}} \right] \,,
\end{align}
\end{subequations}
where $\hat{k}_{j}=k_{j}/|\mathbf{k}|$ and $\mathbf{k}$ is the Fourier wavevector of small and linear fluctuations. The potential $A$ is the perturbation in the lapse function, $B$ the scalar fluctuation in the shift function, and $H_{\rm L}$ and $H_{\rm T}$ are the trace and trace-free scalar perturbations of the $3$-dimensional spatial metric respectively. Our energy-momentum tensor for standard matter components is decomposed as:
\begin{subequations}\label{eq:energy-momentum}
\begin{align} 
	T^{0}_{\phantom{0}0} &= 
	-\sum_\alpha (\rho_{\alpha} + \delta \rho_\alpha) =
	- \sum_\alpha  \rho_\alpha \left( 1  + \delta_\alpha \right) \equiv -  \rho \left( 1  + \delta \right) \,, \\
	T_{{\phantom{0}}0}^i &=  \sum_\alpha ( \rho_\alpha + p_\alpha ) \,\ii \hat k^i v_\alpha \equiv ( \rho + p )\, \ii \hat k^i v \,, \\ 
	T^{i}_{\phantom{i}j} &= \sum_\alpha ( p_\alpha+\delta p_\alpha ) \delta^i_j + \frac{3}{2}(\rho_{\alpha}+p_\alpha) \left( \delta_j^i/3 - \hat k^i \hat k_j \right) \sigma_\alpha \\ \nonumber 
  &\equiv ( p+\delta p ) \delta^i_j + \frac{3}{2}(\rho+p) \left( \delta_j^i/3 - \hat k^i \hat k_j \right)  \sigma  \,,
\end{align}
\end{subequations}
where the index $\alpha$ runs over all matter components, $\delta$ is the matter density contrast, $\delta p$ is the pressure perturbation, $\sigma$ is the anisotropic stress (following Ref.~\cite{MaBert}), and $\rho$ and $p$ are the background density and pressure respectively. Equations (\ref{eq:metric-potentials}) and (\ref{eq:energy-momentum}) are general, and not specialized to any particular coordinate system. 

In previous works, the N-body gauge formulation has been proposed and developed in the context of General Relativity and in k-essence theory. A brief reminder concerning the definition of this specific gauge choice is as follows:

\begin{itemize}
    \item [i)] The temporal slicing is set to the comoving gauge, $B^{\mathrm{Nb}}=v_{\rm m}$. This means that the spatial hypersurfaces are orthogonal to the $4$-velocity of the matter species.
    
    \item[ii)] The spatial threading of the metric is such that $H_{\mathrm{L}}^{\mathrm{Nb}}=0$. This means that at linear order the relativistic density of particles matches the Newtonian one, $\rho_{\mathrm{rel.}} = \rho_{\mathrm{count.}}$. The absence of spatial trace perturbations also implies that $H_{\mathrm{T}}^{\mathrm{Nb}}=3\zeta$, where $\zeta$ is the comoving primordial curvature perturbation.
\end{itemize}
More broadly speaking, this spatial gauge choice can be enforced without the temporal gauge selection, and this is the case for the N-boisson gauge, where the same spatial threading is chosen, but the temporal slicing matches the Poisson gauge one.

The conservation and Euler equations for non-relativistic matter in the N-body gauge are:
\begin{subequations}\label{eq:Nb}
\begin{align}
    & \deltanbprime_{\rm m} +k v_{\rm m}^{\mathrm{Nb}} = 0, \label{eq:consvnb} \\
    & ( \partial_{\tau} + \mathcal{H} ) v_{\rm m}^{\mathrm{Nb}} = -k \left( \Phi +  \gamma^{\text{Nb}}\right) \label{eq:vdivnb},
\end{align}
\end{subequations}
where a prime denotes a derivative with respect to the conformal time, $\tau$,
and they are supplemented by:
\begin{equation}\label{eq:k2gammanb}
    k^{2}\gammanb = -(\partial_{\tau} + \mathcal{H} )\HTnbprime + 12 \pi G_{\rm N} a^{2} \left(\rho+p\right)\sigma,
\end{equation}
and
\begin{equation}\label{eq:poissonnb}
    k^{2}\Phi = 4 \pi G_{\rm N} a^{2} \sum_{\alpha} \delta \rho_{\alpha}^{\textrm{Nb}}.
\end{equation}
The potential $\Phi$ appearing the above equations is the gauge-invariant Bardeen potential, and its definition using the notation given by Equations (\ref{eq:metric-potentials}) is:
\begin{equation}\label{eq:Phi_gen}
    \Phi = H_{\rm L} + \frac{1}{3}H_{\rm T} + \mathcal{H}k^{-1} \left( B - k^{-1}\dot{H}_{\rm T} \right).
\end{equation}
The combination of Equations (\ref{eq:Nb}-\ref{eq:poissonnb}) leads to:
\begin{equation}\label{eq:newteqGR}
    \deltanbprimeprime_{\rm m} + \mathcal{H}\deltanbprime_{\rm m} - 4 \pi G_{\rm N} a^{2} \rho_{\rm m} \deltanb_{\rm m} =  4\pi G_{\rm N} a^{2} \delta \rho_{\text{GR}} ,
\end{equation}
where
\begin{equation}\label{eq:drhoGR}
    \delta \rho_{\text{GR}} = \delta \rho_{\gamma}^{\text{Nb}} + \delta \rho_{\nu}^{\text{Nb}} + \delta \rho_{\text{DE}}^{\text{Nb}} + \delta \rho_{\text{metric}}^{\text{Nb}},
\end{equation}
and we have set
\begin{equation}\label{eq:k2gamma_drho_metrc}
    k^2 \gamma^{\mathrm{Nb}} = 4 \pi G_{\rm N} a^{2} \delta \rho_{\text{metric}} \,.
\end{equation}

As our work is focused on understanding the connection between the N-body gauge and Newtonian simulations, we also present the Newtonian equations of motion:
\begin{subequations}\label{eq:Newton}
\begin{align}  
	 \delta_{\rm m}^{\prime \rm N} + k v_{\rm m}^{\rm N} &= 0 \,,  \label{eq:consvN}\\ 
   \left( \partial_\tau + {\cal H} \right) v_{\rm m}^{\rm N} &= -k \Phi^{\rm N} \,,\label{eq:vdivN}\\
   k^2 \Phi^{\rm N} &= 4\pi G_{\rm N} a^2 \rho_{\rm m} \delta_{\rm m}^{\rm N} \,, \label{eq:poissonN}
\end{align} 
\end{subequations}
which combined give
\begin{equation}\label{eq:newteqN}
    \deltaNprimeprime_{\rm m} + \mathcal{H}\deltaNprime_{\rm m} - 4 \pi G_{\rm N} a^{2} \rho_{\rm m} \deltaN_{\rm m} =  0.
\end{equation}
Equations (\ref{eq:Nb}) and (\ref{eq:Newton}) share the same continuity equation, while the Euler equation would be the same except for the definition of the potential $\Phi$, and the extra relativistic potential $\gammanb$. For a Universe with only non-relativistic matter the former would reduce to the Newtonian potential $\Phi^{\mathrm{N}}$, and the latter would vanish. Thus, the N-body gauge is the natural choice of coordinates that allow us to evolve Newtonian simulations embedded in a relativistic spacetime, in which we can keep track of non-pressureless matter and dark energy perturbations.

\subsection{Horndeski's gravity}\label{sec:Horn}
Horndeski gravity is the most general scalar-tensor theory with second-order differential equations for the metric and the scalar field. Its action is written as:
\begin{equation}
S[g_{\mu\nu},\phi]=\int\mathrm{d}^{4}x\,\sqrt{-g}\left[\sum_{i=2}^{5}\frac{1}{8\pi G_{\rm N}}{\cal L}_{i}[g_{\mu\nu},\phi]\,+\mathcal{L}_{\text{m}}[g_{\mu\nu},\psi_{\rm M}]\right]\,,\label{eq:actionhorn}
\end{equation}
where the $\mathcal{L}_{i}$ terms in the Lagrangian are:
\begin{subequations}
\begin{eqnarray}
{\cal L}_{2} & = & G_{2}(\phi,\,X)\,,\label{eq:L2}\\
{\cal L}_{3} & = & -G_{3}(\phi,\,X)\Box\phi\,,\label{eq:L3}\\
{\cal L}_{4} & = & G_{4}(\phi,\,X)R+G_{4X}(\phi,\,X)\left[\left(\Box\phi\right)^{2}-\phi_{;\mu\nu}\phi^{;\mu\nu}\right]\,,\label{eq:L4}\\
{\cal L}_{5} & = & G_{5}(\phi,\,X)G_{\mu\nu}\phi^{;\mu\nu}-\frac{1}{6}G_{5X}(\phi,\,X)\left[\left(\Box\phi\right)^{3}+2{\phi_{;\mu}}^{\nu}{\phi_{;\nu}}^{\alpha}{\phi_{;\alpha}}^{\mu}-3\phi_{;\mu\nu}\phi^{;\mu\nu}\Box\phi\right] \,. \label{eq:L5}
\end{eqnarray}
\end{subequations}
$X=-\frac{1}{2}\partial_{\mu}\phi\partial^{\mu}\phi$ is the kinetic term of the scalar field, and $\psi_{\rm M}$ represents matter fields minimally coupled to gravity.

Due to general covariance and conservation of the matter energy-momentum tensor, we can consider all Horndeski modifications to the Einstein equations as an effective fluid~\cite{gleyzes,arjona,pace,kk1,defa2}, that is, we simply move all extra coupling between gravity and the scalar field to the right-hand side of the equations, and keep on the left only the original GR terms. In this way the background equations of motion (\ref{eq:actionhorn}) read:
\begin{subequations}
\begin{align}
    H^{2} &= \frac{8 \pi G}{3} \left( \sum_{i}\rho_{i} + \rho_{\mathrm{DE}} \right) \label{eq:H2}\\
    H^{\prime} &= - 4 \pi G a \left[ \sum_{i}\left(\rho_{i} + p_{i}\right) + \rho_{\mathrm{DE}} + p_{\mathrm{DE}}\right]\label{eq:Hprime}
\end{align}
\end{subequations}
where
\begin{subequations}
\begin{align}
\frac{8 \pi G}{3}\mathcal{\rho_{\text{DE}}}\equiv & -\frac{1}{3}G_{2}+\frac{2}{3}X\left(G_{2X}-G_{3\phi}\right)-\frac{2H^{3}\phi^{\prime}X}{3a}\left(7G_{5X}+4XG_{5XX}\right)\label{eq:rhoDE}\\
 & +H^{2}\left[1-\left(1-\alpha_{\textrm{B}}\right)M_{*}^{2}-4X\left(G_{4X}-G_{5\phi}\right)-4X^{2}\left(2G_{4XX}-G_{5\phi X}\right)\right]\nonumber \\
\frac{8 \pi G}{3}p_{\text{DE}}\equiv & \frac{1}{3}G_{2}-\frac{2}{3}X\left(G_{3\phi}-2G_{4\phi\phi}\right)+\frac{4H\phi^{\prime}}{3a}\left(G_{4\phi}-2XG_{4\phi X}+XG_{5\phi\phi}\right)\label{eq:pDE}\\
 & -\frac{\left(\phi^{\prime\prime}-aH\phi^{\prime}\right)}{3\phi^{\prime}a}HM_{*}^{2}\alpha_{\textrm{B}} -\frac{4}{3}H^{2}X^{2}G_{5\phi X}-\left(H^{2}+\frac{2H^{\prime}}{3a}\right)\left(1-M_{*}^{2}\right)\nonumber\\
 &+\frac{2H^{3}\phi^{\prime}XG_{5X}}{3a}\,,\nonumber 
\end{align}
\end{subequations}
with the dark energy background fluid quantities given by:
\begin{subequations}
\begin{align}
    \rho^{\prime}_{\mathrm{DE}} &= -3\mathcal{H}\left(\rho_{\mathrm{DE}}+p_{\mathrm{DE}}\right)\label{eq:rhoprimeDE},\\
    w_{\mathrm{DE}} &= \frac{p_{\mathrm{DE}}}{\rho_{\mathrm{DE}}}\label{eq:wDE}.
\end{align}
\end{subequations}

The equations of motion for linear perturbations are found by varying the action (\ref{eq:actionhorn}) with respect to the synchronous gauge metric:
\begin{equation}\label{eq:synchmetric}
    \dd s^{2} = a(\tau)^2\left[-\dd \tau + \left(\delta_{ij}+h_{ij}\right)\dd x^{i}\dd x^{j}\right],
\end{equation}
where
\begin{equation}\label{eq:synchpots}
    h_{ij}(\mathbf{x},\tau) = \int \dd^{3}k \, e^{\ii \mathbf{k}.\mathbf{x}} \, \left[\hat{k}_i \hat{k}_j h(\mathbf{k,\tau}) + \left(\hat{k}_{i}\hat{k}_j - 1/3 \delta_{ij}\right)6\eta(\mathbf{k},\tau)\right],
\end{equation}
which leads to the following set of equations:
\begin{subequations}
\begin{itemize}
\item Einstein (0,0)
\begin{align}\label{eq:metric_00}
k^{2} \eta - \frac{1}{2} \mathcal{H} h^{\prime} = 4 \pi G_{\rm N} a^{2} \sum_{\alpha} \delta \rho_{\alpha}
\end{align}

\item Einstein (0,i)
\begin{align}\label{eq:metric_0i}
k^{2} \eta = 4 \pi G_{\rm N} a^{2} \sum_{\alpha}\left( \rho_{\alpha} + p_{\alpha} \right) \theta_{\alpha}
\end{align}

\item Einstein (i,j) trace
\begin{align}\label{eq:metric_ii}
h^{\prime \prime} + 2 \mathcal{H} h^{\prime} - 2k^{2}\eta = 8 \pi G_{\rm N} a^{2} \sum_{\alpha} \delta p_{\alpha}
\end{align}

\item Einstein (i,j) traceless
\begin{align}\label{eq:metric_ij}
h^{\prime \prime} + 6\eta^{\prime \prime} + 2\mathcal{H} \left(h^{\prime} + 6\eta^{\prime}\right) - 2k^{2}\eta = - 24 \pi G_{\rm N}a^{2} \sum_{\alpha}\left(\rho_{\alpha} + p_{\alpha}\right)\sigma_{\alpha}.
\end{align}
\end{itemize}
\end{subequations}
Again, the dummy index $\alpha$ runs over all species, dark energy included. These equations are further supplemented by the scalar field perturbation equation:
\begin{align}
D\left(2-\alpha_{\textrm{B}}\right)V_{X}^{\prime\prime}+8aH\lambda_{7}V_{X}^{\prime} & +2a^{2}H^{2}\left[\frac{c_{\text{sN}}^{2}k^{2}}{a^{2}H^{2}}-4\lambda_{8}\right]V_{X}=\frac{2c_{\text{sN}}^{2}}{aH}k^{2}\eta\label{eq:metric_vx}\nonumber \\
 & +\frac{3a}{2HM_{*}^{2}}\left[2\lambda_{2}\delta\rho_{\textrm{tot}}-3\alpha_{\textrm{B}}\left(2-\alpha_{\textrm{B}}\right)\delta p_{\textrm{tot}}\right]\,,
\end{align}
where the subscript ``$\mathrm{tot}$'' refers to all the matter species, not including dark energy. We define each dark energy effective fluid quantity as follows:
\begin{subequations}
\begin{itemize}
    \item Density perturbation:

\begin{equation}\label{eq:drhoDE}
\begin{split}
    \delta \rho_{\mathrm{DE}} &= \delta \rho_{\rm tot} \left(-1-\frac{2}{(\bra-2) \mpl}\right) - \frac{2\bra  }{3 a^2 (\bra-2)}k^2 \eta \\
    &+\frac{2H V_{X}}{3a\mpl\left(\bra-2\right)}\Bigg[ a^2 \Big(H^2 \mpl (3 \bra+\kin)+9 (p_{\rm tot}+\rho_{\rm tot})\Big)\\
    &-3 a \left(\bra-2\right) \mpl H^{\prime}+\bra k^2 \mpl \Bigg] + \frac{3\bra+\kin}{\bra-2}\frac{2H^{2}V_{X}^{\prime}}{3}.
\end{split}
\end{equation}

    \item Velocity divergence:
    
\begin{equation}\label{eq:thetaDE}
\begin{split}
    \left(\rho_{\mathrm{DE}}+p_{\mathrm{DE}}\right)\theta_{\mathrm{DE}} &=  \left[\frac{2 k^2  H'}{3 a}+\frac{1}{3} \alpha_{\textrm{B}} H^2 k^2 +\frac{k^2  (p_{\rm tot}+\rho_{\rm tot})}{M^{2}_{*}}\right]V_{X}\\
    &+\frac{\alpha_{\textrm{B}} H k^2  }{3 a}V_{X}^{\prime}+\theta_{\rm tot} \left(\frac{1}{M^{2}_{*}}-1\right).
\end{split}
\end{equation}

\item Pressure perturbation:
\begin{equation}\label{eq:dpDE}
\begin{split}
    \delta p_{\mathrm{DE}} &= \delta p_{\rm tot} \left(\frac{\kin}{D \mpl}-1\right) - \frac{2 \eta  k^2 (\lambda_{1}-D)}{9 a^2 D} \\
    &- \frac{2 V_{X} \left(3 a^2 H^3 \lambda_{6}+H k^2 \lambda_{5}\right)}{9 a D} - \frac{2 H (D+\lambda_{3}) h'}{9 a D} - \frac{H^2 \lambda_{4} V_{X}^{\prime}}{3 D}.
\end{split}
\end{equation}

    \item Anisotropic stress:
\begin{equation}\label{eq:sigmaDE}
\begin{split}
    \left(\rho_{\mathrm{DE}}+p_{\mathrm{DE}}\right)\sigma_{\mathrm{DE}}&= \frac{ \alpha_{\textrm{M}} H }{9 a}\left(6 \eta^{\prime}+h^{\prime}\right) - \frac{  2  k^2\alpha_{\textrm{T}} }{9 a^2}\eta + \frac{2 H k^2  (\alpha_{\textrm{T}}-\alpha_{\textrm{M}})}{9 a}V_{X}\\
    &-\sigma_{\rm tot} \left(1-\frac{1}{M^{2}_{*}}\right).
\end{split}
\end{equation}
\end{itemize}
\end{subequations}
The $\alpha_{i}$ time-dependent functions that appear in these equations were introduced in~\cite{BS}, and their definition is given in Appendix~\ref{sec:AppA}, along with the $\lambda_{i}$ functions. We also define the scalar field fluctuation as :
\begin{equation}
    V_{X}=a\frac{\delta \phi^{\prime}}{\phi}.
\end{equation}

%For certain theories the correspondence between modified gravity and a fluid is direct, such as the cases of quintessence~\cite{ratra,coble,cald} and k-essence~\cite{arm1,arm2}. However, most scalar-tensor theories, described using the definitions (\ref{eq:drhoDE}--\ref{eq:sigmaDE}), are not endowed with dynamical fluid equations equations of motion, and hence can only be seen as effective fluid quantities. Therefore, 
In the computation of these quantities, we solve the equations for 
synchronous gauge metric potentials plus the scalar field fluctuations, e.g., Equations (\ref{eq:metric_00}-\ref{eq:metric_vx}). After this, the output is rewritten following (\ref{eq:drhoDE}--\ref{eq:sigmaDE}) to obtain the effective fluid quantities.

By choosing to work in this formalism it is possible to introduce the relativistic effects coming from photons, neutrinos (massless and massive) and dark energy directly in the computation of the $\gammanb$ potential, Equation (\ref{eq:k2gammanb}). In order to do this we need to calculate the first and second derivatives of $\HTnb$, as well as the anisotropic stress. The spatial and temporal gauge conditions of the N-body gauge enforce that the traceless perturbation of the spatial metric and the primordial curvature perturbations are related via $\HTnb = 3\zeta$. Combining the (0-i) Einstein equation with the momentum conservation equation~\cite{Nbody1}, both in the N-body gauge, we have that:
\begin{equation}\label{eq:HTnbprime}
        \HTnbprime = 3\frac{\mathcal{H}}{\rho + p} \left[ \left(\rho+p\right)\sigma - \delta p^{\text{S/P}} + p^{\prime}\frac{\theta_{\textrm{tot+DE}}^{\text{S/P}}}{k^{2}} \right],
\end{equation}
and its derivative
\begin{equation}\label{eq:HTnbdotdot}
\begin{split}
     \HTnbprimeprime &= \left[\frac{\mathcal{H}^{\prime}}{\mathcal{H}} - \frac{1}{\left(\rho + p\right)}\left( \rho^{\prime} + p^{\prime}\right)\right]\HTnbprime \\
     &+ 3\frac{\mathcal{H}}{\rho+p}\left[ \left(\rho^{\prime}+p^{\prime}\right)\sigma + \left(\rho+p\right)\sigma^{\prime} - \delta p^{\prime} + p^{\prime\prime}\frac{\theta_{\textrm{tot+DE}}^{\text{S/P}}}{k^{2}} + p^{\prime}\frac{\theta_{\textrm{tot+DE}}^{\prime \text{S/P}}}{k^{2}} \right].
\end{split}
\end{equation}
This is the routine introduced in \hiclass \ to compute both potentials~\cite{hiclass1,hiclass2}.
The photons and neutrinos (massless and massive) pressure and stress perturbations are already available in \hiclass, and to compute their derivatives we have used the Boltzmann equations for the three species, and stored them. For dark energy we compute the pressure and anisotropic stress perturbations (and their derivatives) from Equations (\ref{eq:dpDE}) and (\ref{eq:sigmaDE}). The conformal time derivative of both of these quantities can also be evaluated inside the code, as they are combinations of background quantities\footnote{For the $\lambda_{i}$ functions appearing in the definitions of $\delta p_{\rm DE}$ and $\left( \rho_{\rm DE} + p_{\rm DE} \right)$ we used numerical routines that were already inside the \texttt{background.c} module.} and the synchronous gauge potentials. 
After having computed $\gammanb$ inside the code, we can output the source term $\drhoGR$ in the right-hand side of Equation (\ref{eq:newteqGR}). The other contributions in the definition of $\delta \rho_{\rm GR}$, e.g., the density perturbations of photons, neutrinos and dark energy are evaluated by gauge transforming them from the synchronous gauge to the N-Body gauge, following the prescription:
\begin{equation}
    \delta \rho_{\alpha}^{\rm Nb} = \delta \rho_{\alpha}^{\rm S/P} + 3 \mathcal{H}\rho_{\alpha}\left( 1+ w_{\alpha} \right)\frac{\theta^{\rm S/P}_{\rm tot + DE}}{k^{2}},
\end{equation}
where S and P refers to quantities computed in the synchronous and Poisson gauge respectively, and $\alpha = \gamma,\nu$ and DE. To consistently introduce the relativistic contributions from radiation, neutrinos and dark energy, one can then feed $\drhoGR$ into N-body codes. The simulations, then, are naturally embedded in a relativistic space-time, since the initial displacements and the output of these simulations can be understood in the N-body gauge.

%%%%%%%%%%-----------_%%%%%%%%%%%%%
\begin{table}[h]
    \begin{center} 
        \begin{tabular}{|l| c|} 
            \hline
            Parameter & Value \\
            \hline
            $A_\text{s}$  & $2.215 \times 10^{-9}$ \\
            $n_\text{s}$ & $0.9655$ \\
            $\tau_\text{reio}$ & $0.078$ \\
            $\Omega_{\text{b}}h^{2}$ & $0.02238280$ \\
            $\Omega_{\text{cdm}}h^{2}$ & $0.1201075$ \\
            $\sum m_{\nu}$ & $0.06$ eV \\  
            $h$ & $0.6732$ \\
            \hline						
        \end{tabular}
    \end{center}
    \caption{Cosmological parameter values used in this work.}
    \label{table:class_parameters} 
\end{table}
%%%%%%%%%%-----------_%%%%%%%%%%%%%
Throughout this work we will fix the background expansion of the modified gravity models to be the same as $\Lambda$CDM. The energy densities of photons, massless and massive neutrinos, baryons and cold dark matter, are also fixed to the values in Table~\ref{table:class_parameters}. In this way we are left with four time-dependent functions that characterize Horndeski's theory at linear order, $\bra$, $\run$, $\ten$ and $\kin$, which are defined in terms of the Hordenski functions (see Appendix~\ref{sec:AppA}). The first three will be parametrized as being proportional to the fractional energy density of dark energy, $\alpha_{i} = c_{i} \Omega_{\mathrm{DE}}$, with $i=$B, M and T. The kineticity function, $\kin$, however, will be kept constant in all cases, $\kin=c_{\mathrm{K}}$. Even though we will only work with the parametric form of Horndeski's gravity, e.g., the $\alpha$ function parametrization of Bellini and Sawicki~\cite{BS}, our numerical implementation is $\emph{valid}$ for general covariant theories as well, as presented previously in~\cite{us}.

\begin{figure}[h] 
\centering
\includegraphics[width=0.65\textwidth]{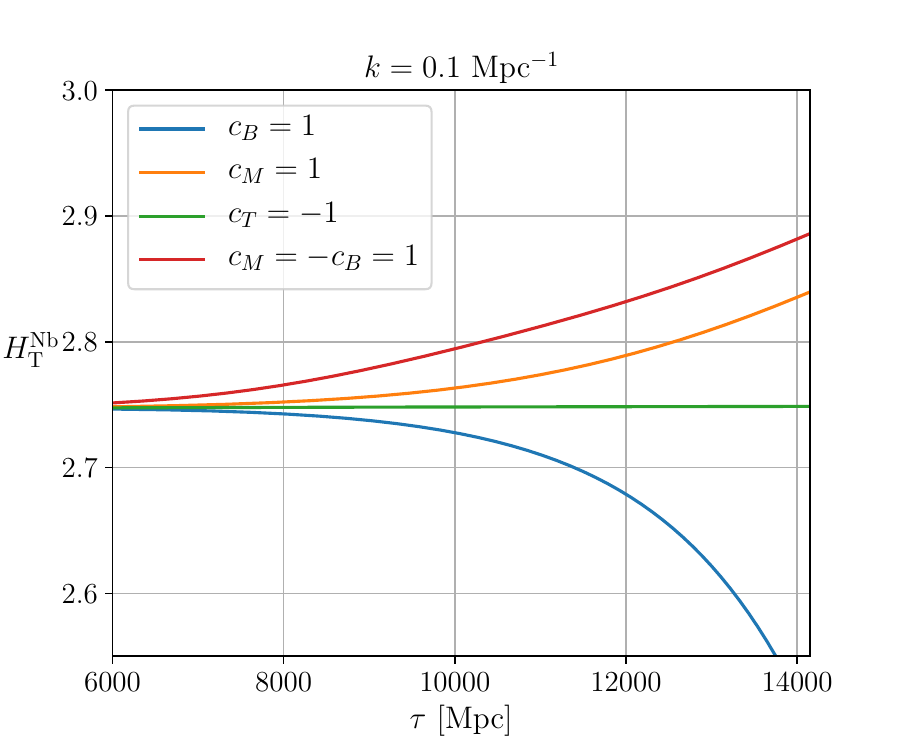} 
\caption{Spatial traceless perturbation in the N-body gauge as a function of conformal time, for a given Fourier mode, for four different modified gravity models, with $\alpha_{i} = c_{i}\Omega_{\mathrm{DE}}$, with fixed $\kin = 1$. Perturbations are normalised so that $\zeta = -1$ on super-horizon scales. 
}
\label{fig:H_T_Nb}
\end{figure}

In Figure~\ref{fig:H_T_Nb} we plot the metric potential $\HTnb$ as a function of conformal time for different models of Horndeski theories. We can see that for different values of the $\alpha_{i}$ functions the metric potential behaves differently.

%%%%%%%%%%-----------_%%%%%%%%%%%%%
%%%%%%%%%%-----------_%%%%%%%%%%%%%

\begin{figure}[h] 
\centering
\includegraphics[width=0.9\textwidth]{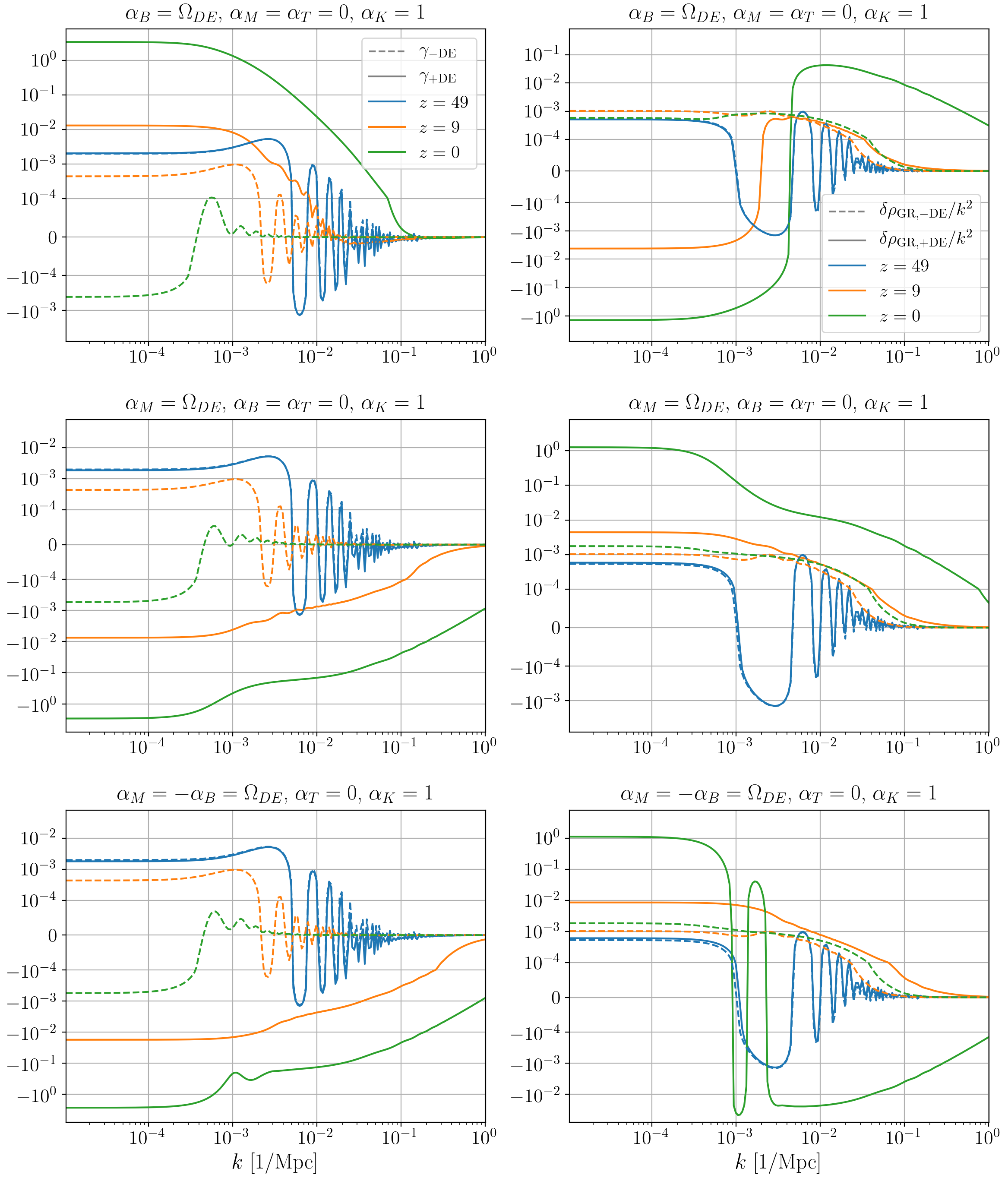} 
\caption{N-body gauge quantities, and the impact of the scalar field on them. \textbf{Left column:} Relativistic potential $\gammanb$ with dark energy perturbations, $\gamma^{\rm Nb}_{\rm +DE}$ (solid line), and without dark energy perturbations, $\gamma^{\rm Nb}_{\rm -DE}$ (dashed line). \textbf{Right column:} Full general relativistic corrections source term, Equation (\ref{eq:drhoGR}), with dark energy perturbations, $\delta \rho_{\mathrm{GR}, \mathrm{+ DE}}$, and without dark energy perturbations,  $\delta \rho_{\mathrm{GR}, \mathrm{- DE}}$. Each row corresponds to a different model of gravity, given in the title of each plot. Blue curves are at redshift $z=49$, orange $z=9$ and green $z=0$. We can see that, in all cases, at early times the two curves overlap, as dark energy perturbations are negligible. At lower redshifts the two are separated, and the solid line gets enhanced at late times. Oscillatory features at intermediate scales are due to photons and massless neutrinos dominating both quantities at early times, but as time goes by they are smoothed out. Perturbations are normalised so that $\zeta = -1$ on super-horizon scales.}
\label{fig:drhoGR_gamma}
\end{figure}

To better understand the behavior and impact of dark energy on the $\gammanb$ potential and the source term, $\drhoGR$, Figure~\ref{fig:drhoGR_gamma} shows the behavior of both of these quantities, with and without scalar field perturbations, in three different models at three different redshifts. The models considered here describe a dark energy field that is only relevant at late times, therefore, at high redshifts the solid and dashed curves overlap. The highly oscillatory behavior of the curves at redshift $z=49$ is due to the fact that the main contributions to $\gammanb$, and $\delta \rho_{\mathrm{GR}}$, are photons and massless neutrinos. At the intermediate redshift of $z=9$, the scalar field perturbations are non-negligible, and the curves get slightly smoothed out. At redshift $z=0$ dark energy perturbations dominate the relativistic potential $\gammanb$ and the source term $\drhoGR$.

%%%%%%%%%%-----------_%%%%%%%%%%%%%
%%%%%%%%%%-----------_%%%%%%%%%%%%%

\section{Impact of relativistic effects}\label{sec:sec3}
The goal of the previous section was to present the theoretical framework for the numerical implementation of $\drhoGR$ in the public code \hiclass. We have performed different consistency checks to ensure that the accuracy of the computation of Equation (\ref{eq:drhoGR}) in the code is well within the desired precision (below $1\%$) for the upcoming stage IV large-scale structure surveys. Additionally, several state-of-the-art modified gravity Einstein-Boltzmann solvers have been compared in~\cite{mg_eb}, with overall agreement below the percent-level threshold as well. In this section we move to present and discuss the behavior and impact of the dark energy scalar field relativistic effects in the linear Newtonian matter power spectrum. We focus only on the regime in which linear perturbation theory is valid, where the N-body gauge is known to be safe to use. 

Our method is intended to be used in combination with Newtonian N-body simulations, which accurately capture non-linear dynamics. The relativistic effects including the contribution coming from dark energy can be added to Newtonian simulations by implementing the linear density field, $\delta \rho_{\rm GR}$. This approach has been used to perform N-body simulations that are fully compatible with GR including the effect of linear dark energy perturbations as well as massive neutrinos \cite{Nbody5, Dakin:2019vnj, Knabenhans:2020gdo}. Using the effective fluid approach, we can perform Newtonian N-body simulations that are fully compatible with Horndeski gravity on linear scales without additional computational cost. In the case of Horndeski gravity, care must be taken about the linearity of $\delta \rho_{\rm GR}$. In general, this density field contains a contribution from the matter density field $\delta \rho_{\rm m}$, which becomes non-linear on small scales. We will present a method to separate this contribution from $\delta \rho_{\rm GR}$ and include it in Newtonian simulations as a time-dependent effective gravitational constant. In this way, the linearity of $\delta \rho_{\rm GR}$ is ensured even on small scales.

%, and as discussed before, can be naturally embedded in a relativistic space-time using the N-body gauge prescription. As has already been shown in~\cite{Nbody1}, the relativistic effects of photons and neutrinos are subdominant at small scales. Therefore, we will focus only on the contributions coming from dark energy in the power spectrum. 

\subsection{General description}\label{sec:MG_gen}

Modified gravity affects the matter power spectrum in different ways across all scales. In this section we will focus only on the largest scale effects, $k \sim 10^{-5}-10^{-2}$ Mpc$^{-1}$, and leave to the next section a more in-depth discussion of the imprint that dark energy leaves in other regimes. The kineticity function is related to the kinetic energy of the scalar field perturbations, and therefore it is only relevant at scales near the horizon. The other three functions, however, affect the power spectrum on all scales. Therefore, we will study the impact of these three functions in the power spectrum for several combinations and vary the kineticity to see its effect on large scales.

%%%%%%%%%%-----------_%%%%%%%%%%%%%
\begin{figure}[h] 
\centering
\includegraphics[width=1.\textwidth]{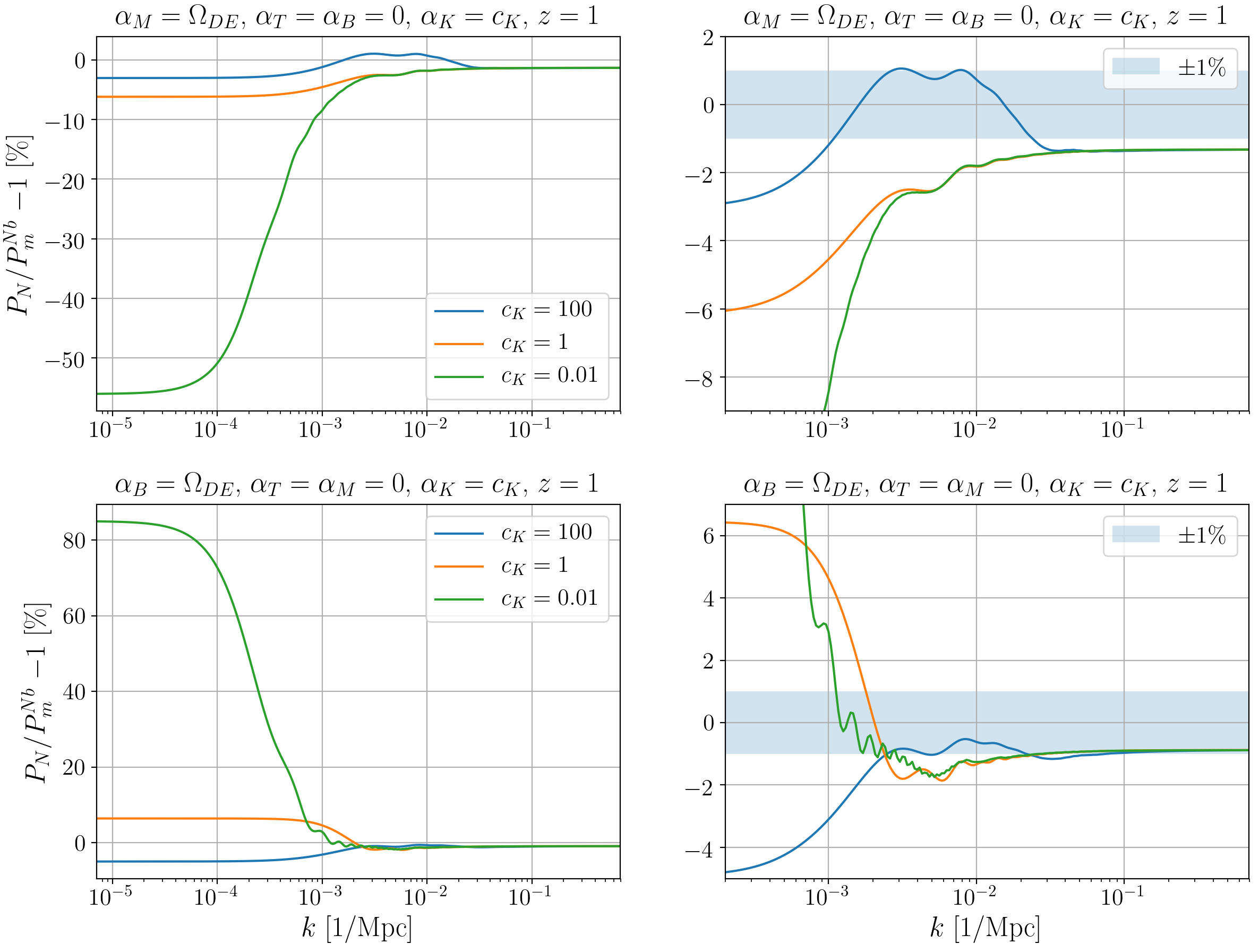} 
\caption{Relative difference between the linear Newtonian matter power spectrum $P_{\rm N}$, Equation (\ref{eq:newteqN}), and 
the matter power spectrum in N-body gauge in Horndeski gravity $P^{\rm Nb}_{\rm m}$, Equation (\ref{eq:newteqGR}). The top row corresponds to a modified gravity model in which we only have the running of the Planck mass, $\run$, and the bottom row corresponds to just braiding, $\bra$. On the left hand side plots, we show the full interval in $k$ while on the right we show only the scales probed by future LSS stage IV surveys. In all cases, we show three different kineticity functions, $c_{\rm K}=100, 1, 0.01$. We also plot the $1\%$ deviation region (shaded blue). Our initial conditions for $\delta_{\rm m}^{\rm Nb}$ are set at $a= 0.01$ ($z= 99$).}%We can see the enhancement of the amplitude at large scales caused by the kineticity in both cases.}
\label{fig:plot1}
\end{figure}
%%%%%%%%%%-----------_%%%%%%%%%%%%%
%%%%%%%%%%-----------_%%%%%%%%%%%%%
\begin{figure}[h] 
\centering
\includegraphics[width=1.\textwidth]{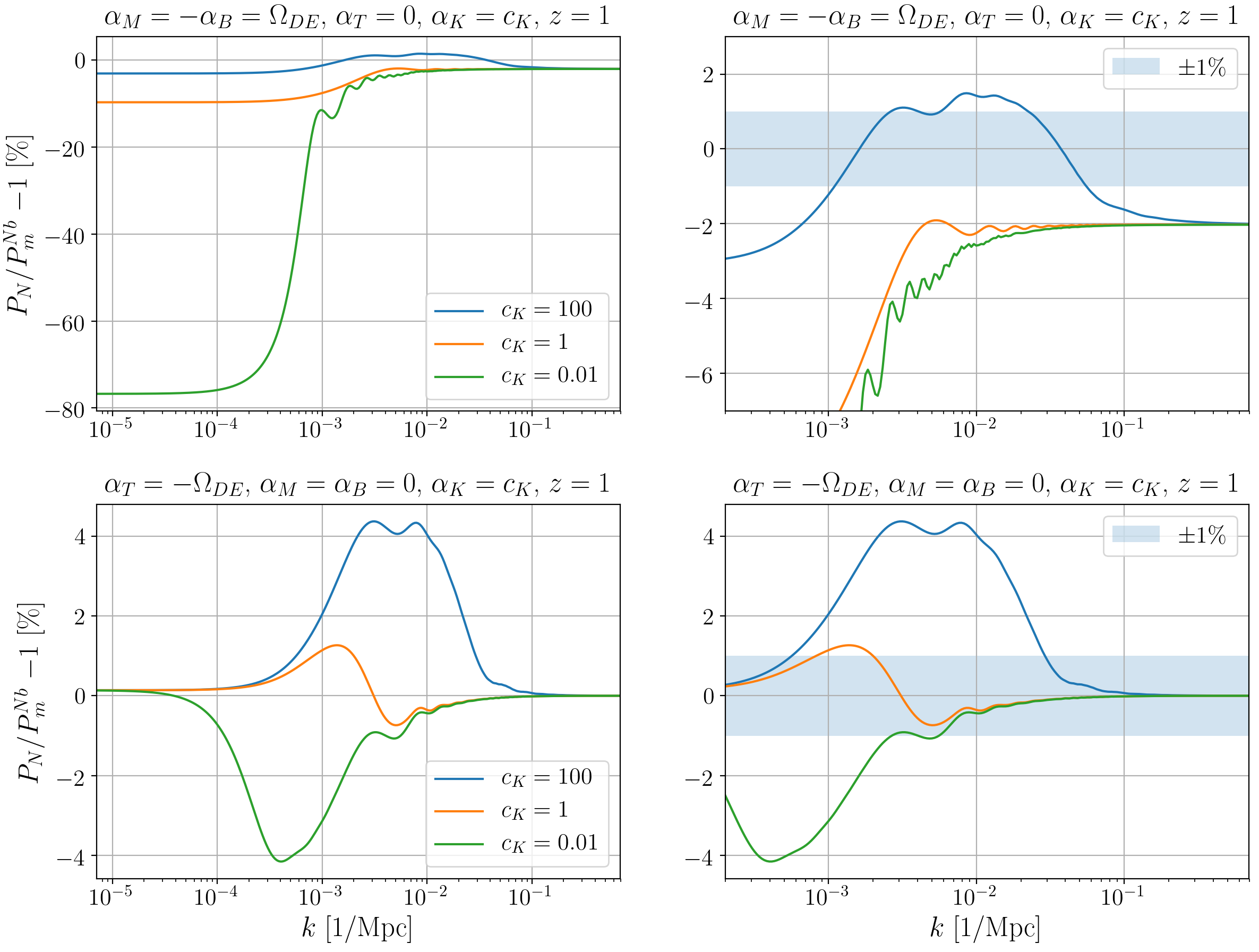} 
\caption{Relative difference between the linear Newtonian matter power spectrum $P_{\rm N}$, Equation (\ref{eq:newteqN}), and the matter power spectrum in N-body gauge in Horndeski gravity $P^{\rm Nb}_{\rm m}$, Equation (\ref{eq:newteqGR}). The top row corresponds to a modified gravity model with a Jordan-Brans-Dicke (JBD) parametrization, $\run=-\bra$, and the bottom row shows the case only with the tensor modification, $\ten$. On the left hand side plots, we show the full interval in $k$, and on the right we show the interval of scales probed by future LSS stage IV surveys. In all cases we show three different kineticity functions, $c_{\rm K}=100, 1, 0.01$. We also plot the $1\%$ deviation region (shaded blue). Our initial conditions for $\delta_{\rm m}^{\rm Nb}$ are set at $a= 0.01$ ($z= 99$).}% Contrary to Figure~\ref{fig:plot1}, the kineticity does not affect the behavior at large scales in the case with only the tensor modification.}
\label{fig:plot2}
\end{figure}
%%%%%%%%%%-----------_%%%%%%%%%%%%%

We choose to assign three different values to $c_{\mathrm{K}}$: $100$, $1$ and $0.01$, while the other constants of proportionality between the $\alpha$ functions and the fractional energy density of dark energy will be kept fixed. In Figures \ref{fig:plot1} and \ref{fig:plot2}, we present the relative difference between the Newtonian linear matter power spectrum $P_{\rm N}$, which is the solution of Equation (\ref{eq:newteqN}), and the matter power spectrum in N-body gauge in Horndeki gravity, $P_{\rm m}^{\rm Nb}$, which is the solution of Equation (\ref{eq:newteqGR}). Each row corresponds to a given model, with the plots on the left showing the full range of length scales, and the ones on the right showing a smaller range. There are multiple constraints on $c_{\mathrm{B}}$, $c_{\mathrm{M}}$ and $c_{\mathrm{T}}$ from different cosmological datasets. The tightest one, for instance, sets $|c_{\mathrm{T}}|<\mathcal{O}(10^{-15})$~\cite{gw_grb}, while the others must be $< \mathcal{O}(10^{-2})$ for $c_{\mathrm{B}}$~\cite{cremi}, and $< \mathcal{O}(10^{-3})$ for $c_{\mathrm{M}}$~\cite{burr}. However, we have chosen to fix all of them to $1$ to better highlight their effect on the matter power spectrum. Another important aspect to consider is the relation between the sound speed with which scalar field fluctuations propagate, and the kineticity. In Horndeski theories, the sound speed is given by:
\begin{equation}
    c_{\text{s}}^{2}=  \frac{1}{\alpha_{\rm K} + \frac{3}{2}\bra^{2}}\left[\left(2-\alpha_{\textrm{B}}\right)\left(-\frac{H^{\prime}}{aH^{2}}+\frac{1}{2}\alpha_{\textrm{B}}\left(1+\alpha_{\textrm{T}}\right)+\alpha_{\textrm{M}}-\alpha_{\textrm{T}}\right)-\frac{3\left(\rho_{\textrm{tot}}+p_{\textrm{tot}}\right)}{H^{2}M_{*}^{2}}+\frac{\alpha_{\textrm{B}}^{\prime}}{aH}\right].\label{eq:cs2}
\end{equation}
By inspecting this equation we can see that the kineticity and sound speed are inversely proportional to one another (as long as $\bra^{2} \ll \kin$), in such a way that the larger the value of $c_{\mathrm{K}}$, the smaller $c_{\rm s}^{2}$, and vice-versa.

\begin{figure}[h]
\centering
\includegraphics[width=1.\textwidth]{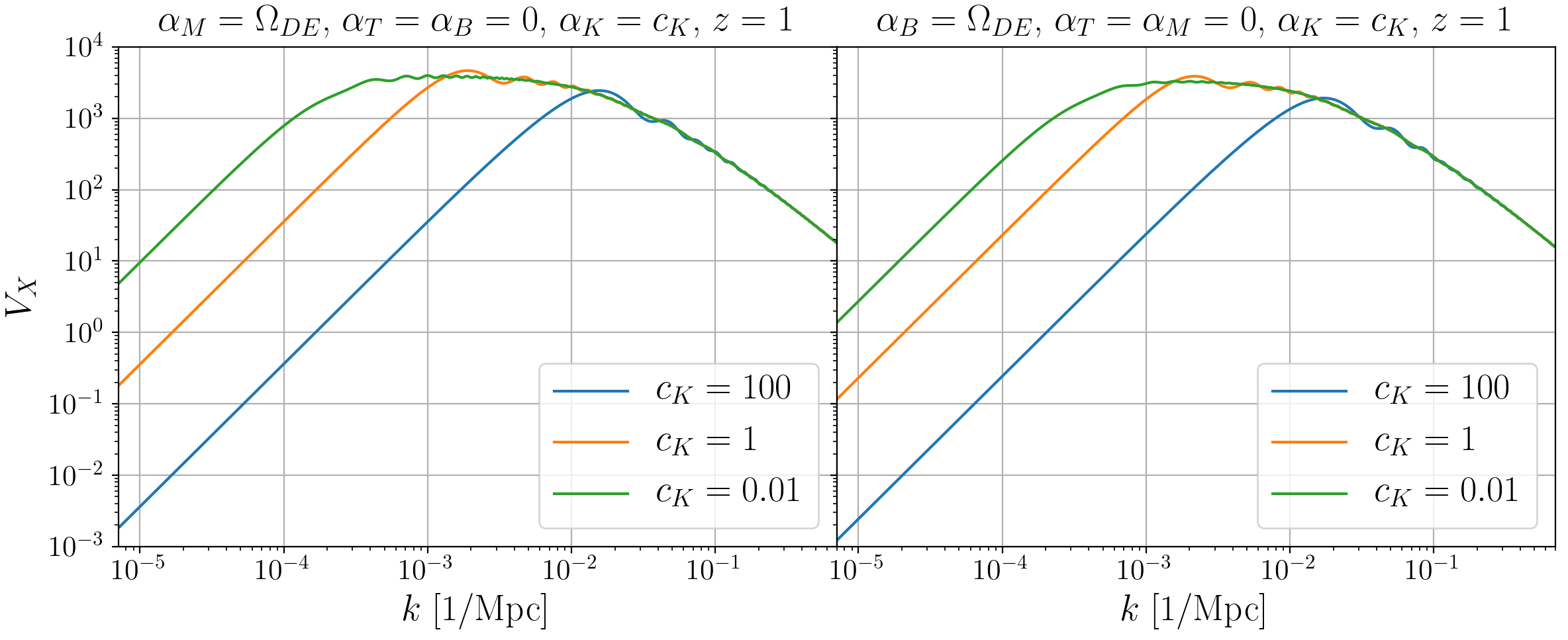} 
\caption{Scalar field fluctuation, $V_{X}$, as a function of scale for two different gravity models: $\kin = c_{\rm K}$ and $\run = \Omega_{\rm DE}$, $\kin = c_{\rm K}$ and $\bra = \Omega_{\rm DE}$ % only running and only braiding
, at fixed redshift, $z=1$. We vary the kineticity in three different constant values, $c_{\rm K} = 100$, $1$ and $0.01$. We can see that at large scales, smaller values of $c_{\rm K}$ have larger amplitudes. This behavior causes the enhancing of the signal at small $k$ seen in Figures~\ref{fig:plot1} and \ref{fig:plot2}.}
\label{fig:V_x_plots}
\end{figure}

The plots in Figures \ref{fig:plot1} and \ref{fig:plot2} show that the signal is greater at small wavenumbers, reaching up to above $80\%$ for the case with just the braiding being non-zero (bottom row of Figure \ref{fig:plot1}). The reason for this large amplitude at these scales can be understood by analyzing the scalar field fluctuation equation, (\ref{eq:metric_vx}). The dominant term at large scales in this equation is proportional to the synchronous gauge metric potential $\eta$ times the sound speed. Since $\eta$ does not depend on the kineticity -- Equation (\ref{eq:metric_0i}) is independent of $\kin$ -- it is the same for different values of $c_{\mathrm{K}}$ (remember our background is fixed for a $\Lambda$CDM background). As we have seen, $c_{\rm s}^{2}$ is inversely proportional to $\kin$, which increases considerably the amplitude of the scalar field fluctuations, as seen in Figure~\ref{fig:V_x_plots}, and consequentially affects the value at large scales of the relativistic correction.
%%%%%%%%%%-----------_%%%%%%%%%%%%%
%%%%%%%%%%-----------_%%%%%%%%%%%%%
We will now discuss the behavior at intermediate scales for each model individually. When there is only braiding (bottom row of Figure \ref{fig:plot1}), at smaller $k$ there is also a dependence on the value of the sound speed, as for $c_{\mathrm{K}}=100$, we have an enhancement of the relativistic power spectrum $P^{\rm Nb}_{\rm m}$ with respect to the linear Newtonian one. For smaller values of the kineticity, however, after starting enhancing the power spectrum we see that there is a shift to the opposite direction, towards suppressing the power spectrum at intermediate to large scales.

This behavior is absent in the running-only case (top row of Figure~\ref{fig:plot1}), where the sign of the relative difference at large scales is independent of the sound speed. For the mixed case, Jordan-Brans-Dicke parametrization $\alpha_{\mathrm{M}}=-\alpha_{\mathrm{B}}$, the same independent of $\kin$ behavior is present. Unlike the former cases, however, the model in which there are only modifications of gravity in the tensor sector, $\ten \neq 0$, the bulk of the modified gravity signal is not at small $k$, but at intermediate scales, where all the curves show a bump in the signal, and then show no enhancement or suppression at large scales. Although not shown in this work, if the opposite sign of the $c_{\mathrm{M}}$, $c_{\mathrm{B}}$ and $c_{\mathrm{T}}$ constants is chosen, the effect on the power spectrum becomes opposite.

The final range of wavenumbers is at small scales, e.g., $k>10^{-1}$ Mpc$^{-1}$. In this range, in all but the $\ten \neq 0$ case, there is a vertical shift with respect to the $0\%$ line. This displacement is caused by the gravitational constant, $G_{\rm eff}$, which in scalar-tensor theories, in general, is not constant, and becomes time-dependent. This effective gravitational constant is scale independent at linear order when we take $k \to \infty$, and can be computed and implemented in modified gravity Newtonian simulations. A more detailed discussion on this topic will be carried out in the next section.

\subsection{Separating small-scale effects from relativistic effects}\label{sec:MG_QSA}

It is well known that for large values of $k$, gravity can be described using Newtonian gravity, a fundamental pillar of Newtonian N-body codes. The set of equations these simulations are solving is the discretized phase-space equivalent of Equations (\ref{eq:consvN}-\ref{eq:vdivN}) and (\ref{eq:poissonN}), for a colisionless cold dark matter fluid, the so-called Vlasov-Poisson equations. To incorporate modified gravity effects in these codes, the Newtonian gravitational constant $G_{\rm N}$ is replaced with an effective gravitational constant $G_{\rm eff}$, that captures the small scale effects of the scalar field. 

\begin{figure}[h] 
\centering
\includegraphics[width=1.\textwidth]{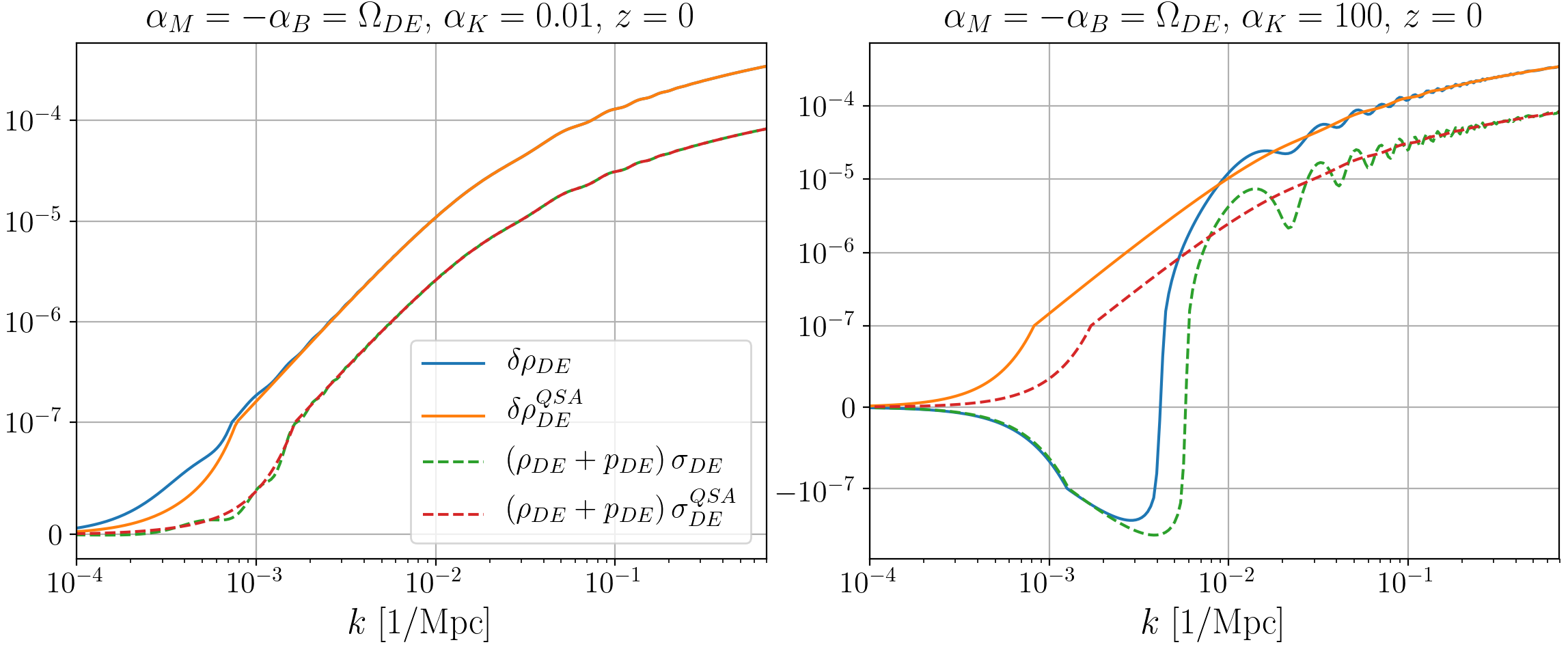} 
\caption{Comparison of the full dark energy density perturbation (solid lines) and anisotropic stress (dashed lines), and their QSA counterparts, as a function of scale at a fixed redshift, $z=0$. On the left plot we show the evolution for a small kineticity value, $c_{\rm K}=0.01$, and on the right a larger value, $c_{\rm K}=100$. For $c_{\rm K}=0.01$, the full and QSA contributions overlap at smaller values of $k$, while for $c_{\rm K}=100$ this happens only at large values of $k$. This is due to the sound speed of the scalar field, which is larger for $c_{\rm K}=0.01$ than it is for $c_{\rm K}=100$, which pushes the QSA regime of validity to smaller scales.}
\label{fig:DE_QSA}
\end{figure}

The specific form of $G_{\rm eff}$ for Horndeski theory is found by taking the quasi-static approximation (QSA)~\cite{qsa_BS,qsa_pace} in the Einstein field equations and the scalar field fluctuation equation. This QSA routine considers that the evolution time-scale of the perturbation of $\phi$ is much smaller than the Hubble rate, therefore, specific time derivatives in the equations of motion may be safely neglected. In the synchronous gauge used in Equations (\ref{eq:drhoDE}-\ref{eq:sigmaDE}), we will use the following prescription to implement the QSA:
\begin{itemize}
    \item We neglect the following terms in the Einstein and scalar field equations: $\eta^{\prime}$, $\eta^{\prime\prime}$, $V_{X}^{\prime}$, $V_{X}^{\prime\prime}$, $\delta p_{\rm tot}$, $\sigma_{\rm tot}$.
    \item We then have algebraic relations between the perturbations in the matter density, $\delta \rho_{\rm tot}$, and the remaining metric potentials and scalar field fluctuation.
    \item We next substitute these relations into Equations (\ref{eq:drhoDE}-\ref{eq:sigmaDE}).
\end{itemize}
By doing so one can separate $\delta \rho_{{\rm DE}}$ into two parts:
\begin{equation}\label{eq:drhoDE_decomp}
    \delta \rho_{\mathrm{DE}} = \delta \rho_{\mathrm{DE}}^{\mathrm{QSA}} + \delta \rho_{\mathrm{DE}, \ \mathrm{rel.}},
\end{equation}
where $\delta \rho_{\mathrm{DE}}^{\mathrm{QSA}}$ is the QSA contribution to the dark energy density perturbation, and $\delta \rho_{\mathrm{DE}, \ \mathrm{rel.}}$ encapsulates all the other terms that are not proportional to matter density perturbations. The same procedure must also be performed for $\sigma_{\rm DE}$, as this quantity captures the amount in which the Poisson gauge potentials differ from each other (in GR they are the same). Thus, we separate the anisotropic stress as:
\begin{equation}\label{eq:sigmaDE_decomp}
    \sigma_{\mathrm{DE}}= \sigma_{\mathrm{DE}}^{\mathrm{QSA}} + \sigma_{\mathrm{DE}, \ \mathrm{rel.}},
\end{equation}
where the quantities with the superscript QSA refer to the QSA contribution. Following the scheme outlined above, we find:
\begin{align}
    \delta \rho_{\mathrm{DE}}^{\mathrm{QSA}} &= \left( \frac{\bra \lambda_{2} - \csN\left[ 2 + M_{*}^{2}\left(\bra -2\right) \right]}{\csN \mpl \left( \bra -2 \right)} \right) \delta \rho_{\rm m}, \label{eq:drhoDE_QSA}\\
    \left(\rho_{\mathrm{DE}} + p_{\mathrm{DE}} \right)\sigma_{\mathrm{DE}}^{\mathrm{QSA}} &= \left( \frac{\run \left[ \bra + 2\run + \ten\left(\bra - 2\right) \right] + \ten \lambda_{2}}{3 \csN \mpl}\right) \delta \rho_{\rm m}. \label{eq:sigmaDE_QSA}
\end{align}
Figure~\ref{fig:DE_QSA} shows the behavior of the full dark energy density perturbations and anisotropic stress at $z=0$, and their QSA counterparts. We can see that the full perturbations and the QSA contribution overlap when we move to larger $k$ values, as expected.
%%%%%%%%%%-----------_%%%%%%%%%%%%%
%%%%%%%%%%-----------_%%%%%%%%%%%%%
After substituting Equation (\ref{eq:drhoDE_decomp}) and (\ref{eq:sigmaDE_decomp}) into Equation (\ref{eq:newteqGR}), we move the terms proportional to $\delta \rho_{\rm m}$ to the left hand side, so we can rewrite Equation (\ref{eq:newteqGR}) as:
\begin{equation}\label{eq:newteqGR_Geff}
    \deltanbprimeprime_{\rm m} + \mathcal{H}\deltanbprime_{\rm m} - 4 \pi G_{\rm eff} a^{2} \rho_{\rm m} \deltanb_{\rm m} =  4\pi G_{\rm N} a^{2} \delta \rho_{\mathrm{GR}, \ \mathrm{rel.}},
\end{equation}
with
\begin{align}
    \delta \rho_{\mathrm{GR}, \ \mathrm{rel.}} &= \delta \rho_{\gamma}^{\text{Nb}} + \delta \rho_{\nu}^{\text{Nb}} + \delta \rho_{\mathrm{DE}, \ \mathrm{rel.}}^{\text{Nb}}  + \delta \rho_{\text{metric, rel.}}, \\
    \delta \rho_{\mathrm{DE}, \ \mathrm{rel.}}^{\text{Nb}} &= \delta \rho_{\mathrm{DE}}^{\text{Nb}} - \delta \rho_{\mathrm{DE}}^{\mathrm{QSA}}, \\
    k^{2} \gammanb_{\mathrm{rel.}} &= 4 \pi G \delta \rho_{\mathrm{metric, \ rel.}},
\end{align}
and the effective gravitational constant given by
\begin{equation}\label{eq:Geff}
    G_{\rm eff} = 1 + \frac{ c_{\mathrm{sN}}^{2} \left( 2- 2M_{*}^{2} + 2\alpha_{\mathrm{T}} \right) + \left( \alpha_{\mathrm{B}} + 2 \alpha_{\mathrm{M}} -2\alpha_{\mathrm{T}} + \alpha_{\mathrm{B}}\alpha_{\mathrm{T}} \right)^{2}}{2c_{\mathrm{sN}}^{2} M_{*}^{2}}.
\end{equation}
In conventional modified gravity Newtonian simulations~\cite{mg_nb,mg_cola} the right hand side of equation (\ref{eq:newteqGR_Geff}) is absent, and the codes are solving the usual cold dark matter fluid equation using Newtonian gravity, with the Newtonian potential $\Phi^{\rm N}$ given by:
\begin{equation}
    \nabla^{2} \Phi^{\rm N} = 4 \pi a^{2} G_{\rm eff} \rho_{\rm m} \delta_{\rm m}^{\rm N},
\end{equation}
and the evolution equation is then
\begin{equation}\label{eq:newteq_Geff}
    \deltanbprimeprime_{\rm m} + \mathcal{H}\deltanbprime_{\rm m} - 4 \pi G_{\rm eff} a^{2} \rho_{\rm m} \deltanb_{\rm m} =  0.
\end{equation}
The difference from our method to incorporate relativistic effects coming not only from non-presureless matter species, but also from the Horndeski scalar field, are the terms in $\delta \rho_{\mathrm{GR}, \ \mathrm{rel.}}$ and $\sigma_{\rm DE, \ \mathrm{rel.}}$. Formally speaking, the solution of Equation (\ref{eq:newteqGR_Geff}) is the same as the solution from Equation (\ref{eq:newteqGR}), since it is just a recasting of the same equation. This fact is what allows us to quantify exactly the effects coming solely from relativistic corrections introduced by modified gravity on large scales. Hence, by comparing the matter power spectrum built from the solution on Equation (\ref{eq:newteq_Geff}) with respect to the one built from (\ref{eq:newteqGR_Geff}), the effects introduced by $G_{\rm eff}$ are mitigated, as the homogeneous solution of both of these equations is the same.
%%%%%%%%%%-----------_%%%%%%%%%%%%%
\begin{figure}[h] 
\centering

\includegraphics[width=1.\textwidth]{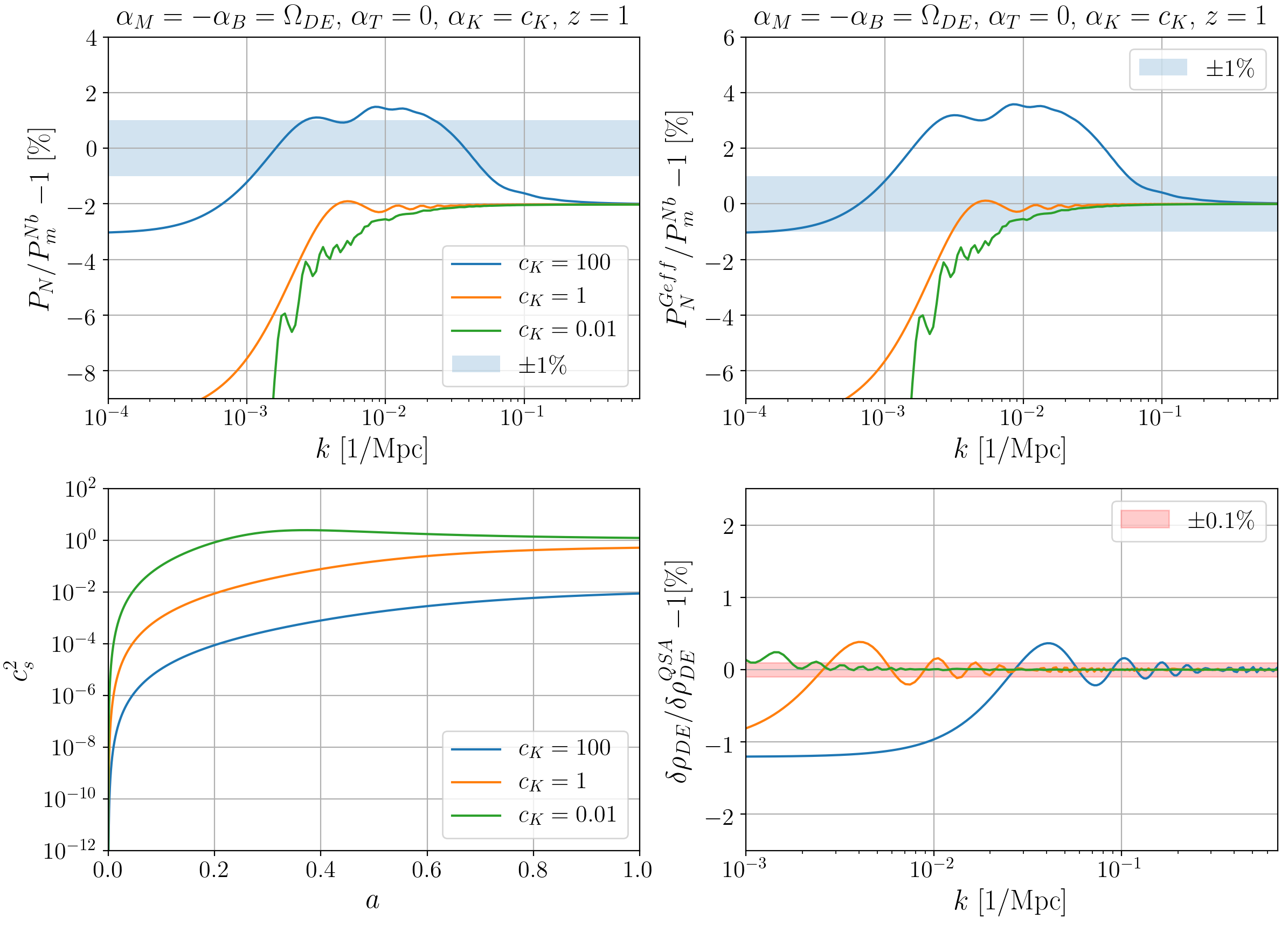} 
\caption{Impact of separating modified gravity effects on smalls scales in the matter power spectrum. \textbf{Top left:} Relative difference between the linear Newtonian matter power spectrum described by Equation (\ref{eq:newteqN}), and the N-body gauge matter power spectrum in Horndeksi gravity at redshift $z=1$. \textbf{Top right:} Relative difference between the linear Newtonian matter power spectrum with $G_{\rm eff}$, $P_{\rm N}^{G_{\rm eff}}$, described by Equation (\ref{eq:newteq_Geff}), and the N-body gauge matter spectrum in Horndeski gravity at redshift $z=1$. 
\textbf{Bottom left:} Squared sound speed of the scalar field as a function of the scale factor. \textbf{Bottom right:} Relative difference between the full dark energy density perturbation, Equation (\ref{eq:drhoDE}), and its QSA counterpart, Equation (\ref{eq:drhoDE_QSA}). All plots are for the same gravity model, Jordan-Brans-Dicke parametrization, and we only vary the values of the kineticity, $c_{\rm K} =100, 1, 0.01$. The top right plot shows the $1\%$ deviation region (shaded blue) in which we can see that purely relativistic effects are not captured by $G_{\rm eff}$, and exceed the percent-level deviation threshold at scales probed by future LSS stage IV surveys. In the bottom right plot, we show the $0.1\%$ deviation interval (shaded red), where the relativistic contribution decay when we move to larger values of $k$, thus, ensuring a smooth transition to the regime of Newtonian gravity. This exhibits the validity of our formalism to implement relativistic effects in Newtonian N-body simulations. Our initial conditions for $\delta_{\rm m}^{\rm Nb}$ are set at $a= 0.01$ ($z= 99$).}
\label{fig:JBD_plot}
\end{figure}
%%%%%%%%%%-----------_%%%%%%%%%%%%%

\begin{figure}[h] 
\centering
\includegraphics[width=1.\textwidth]{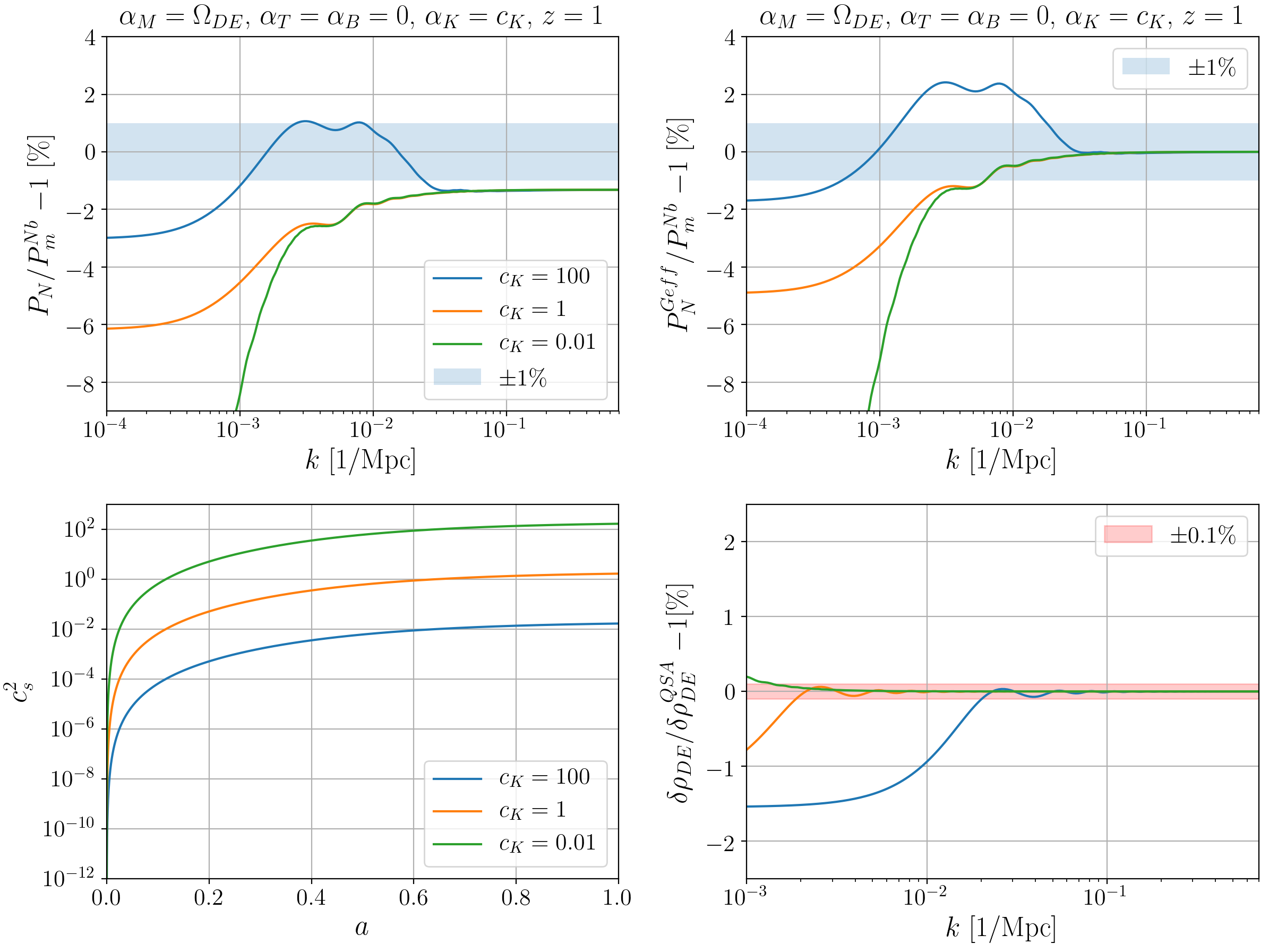} 
\caption{Same as Figure~\ref{fig:JBD_plot} but for a gravity model where modifications are characterized only by the running of the Planck mass, $\run$.}
\label{fig:cm_plot}
\end{figure}

Figures \ref{fig:JBD_plot} and \ref{fig:cm_plot} show the separation of these two effects and the impact of relativistic corrections in two models at $z=1$. The top panels show the relative difference in percentage between the linear Newtonian matter power spectrum with and without $G_{\rm eff}$ effects, $P_{\rm N}^{\rm Geff}$ (solution of \ref{eq:newteq_Geff}) and $P_{\rm N}$ (solution of \ref{eq:newteqN}) respectively, and the N-body gauge matter power spectrum, in Horndeski gravity, $P^{\rm Nb}_{\rm m}$ (solution of \ref{eq:newteqGR_Geff}). The bottom panels show the square of the sound speed of the scalar field and the relative difference between the energy density perturbations of dark energy and its QSA counterpart. As modified gravity Newtonian simulations use the QSA limit, a good check to see if our formalism will have a smooth transition from linear perturbation theory to Newtonian gravity is to quantify the agreement between the full and the QSA contribution to dark energy density perturbations. We chose to present models that have a below $0.1 \%$ agreement between $\delta \rho_{\mathrm{DE}}^{\text{Nb}}$ and $\delta \rho_{\mathrm{DE}}^{\mathrm{QSA}}$, at scales $k \gtrapprox 0.1$ Mpc$^{-1}$. This is roughly the scale at which linear theory breaks, and where the Newtonian approximation is correctly describing gravity. 
%%%%%%%%%%-----------_%%%%%%%%%%%%%
%%%%%%%%%%-----------_%%%%%%%%%%%%%
The top right plots of Figures \ref{fig:JBD_plot} and \ref{fig:cm_plot} show the effects coming purely from relativistic effects of modified gravity. The deviations between both spectra may be above the $1 \%$ level, the usual required accuracy in these simulations. This shows that in order to make consistent simulations in modified gravity we must include the relativistic source term, $\delta \rho_{\mathrm{GR}}$, in simulations.

\subsection{Gravity acoustic oscillations}
%Oscillatory features
\label{sec:MG_osc}

In the figures presented in the previous subsections, we see the emergence of oscillatory features in the range $k \sim 10^{-3} - 10^{-2}$ Mpc$^{-1}$. In this section we will investigate these oscillations more closely. 

\begin{figure}[h] 
\centering
\includegraphics[width=1.\textwidth]{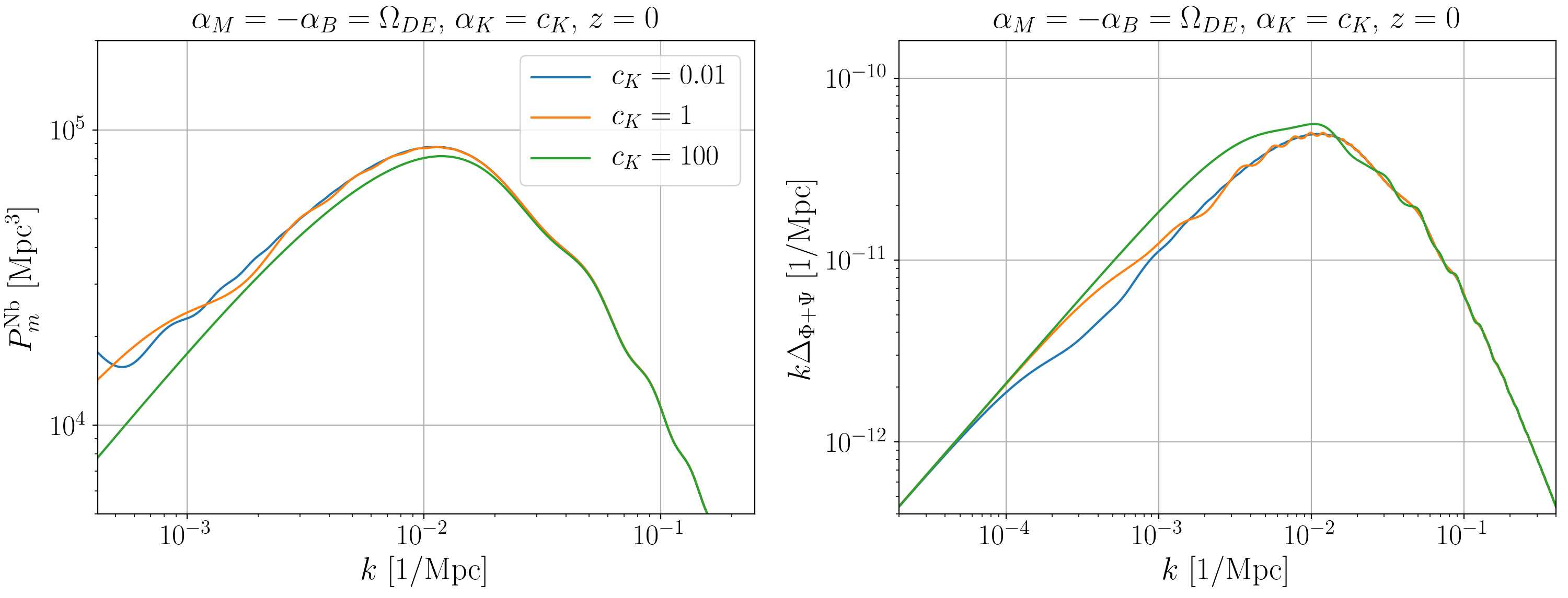} 
\caption{Oscillatory features in observable quantities. \textbf{Left:} N-body gauge matter power spectrum at redshift $z=0.$. \textbf{Right:} Lensing potential transfer function for the same model. Gravity acoustic oscillations appear in the matter power spectrum only in the model with low values of $c_{\rm K}$ at scales below the scale of matter-radiation equality, $k_{\rm eq.} \sim 10^{-2}$ Mpc$^{-1}$. Since the parametrization is chosen to be proportional to the fractional dark energy density, $\Omega_{\rm DE}$, modified gravity effects will only affect the matter power spectrum during the late stages of matter domination. Therefore these features do not affect the BAO oscillations, where the QSA contribution is already dominating the dark energy density perturbations, thus damping the GAOs. The same smooth behavior at small $k$ values for lower values of the kineticity is also present in the lensing potential transfer function. However, due to the presence of anisotropic stress (gravitational slip in the Newtonian gauge potentials), for $c_{\rm K}=100$, the oscillations are also present at values of $k$ bigger than $k_{\rm eq.}$ in the lensing potential.}
\label{fig:mpk_osc}
\end{figure}

From the scales in which these oscillations appear, and allied with the fact that QSA contributions do not oscillate since dynamical equations become constraint equations in the QSA, we can identify these features as purely relativistic effects of modified gravity, and they reveal directly the dynamical nature of the additional degree of freedom.  Figure~\ref{fig:mpk_osc} shows the matter power spectrum in the N-body gauge on the left, and the lensing potential on the right. We can see that oscillations are present in both of these observables, and therefore can be probed by future LSS and 21cm intensity mapping surveys.
%%%%%%%%%%-----------_%%%%%%%%%%%%%
%%%%%%%%%%-----------_%%%%%%%%%%%%%

\begin{figure}[h] 
\centering
\includegraphics[width=1.\textwidth]{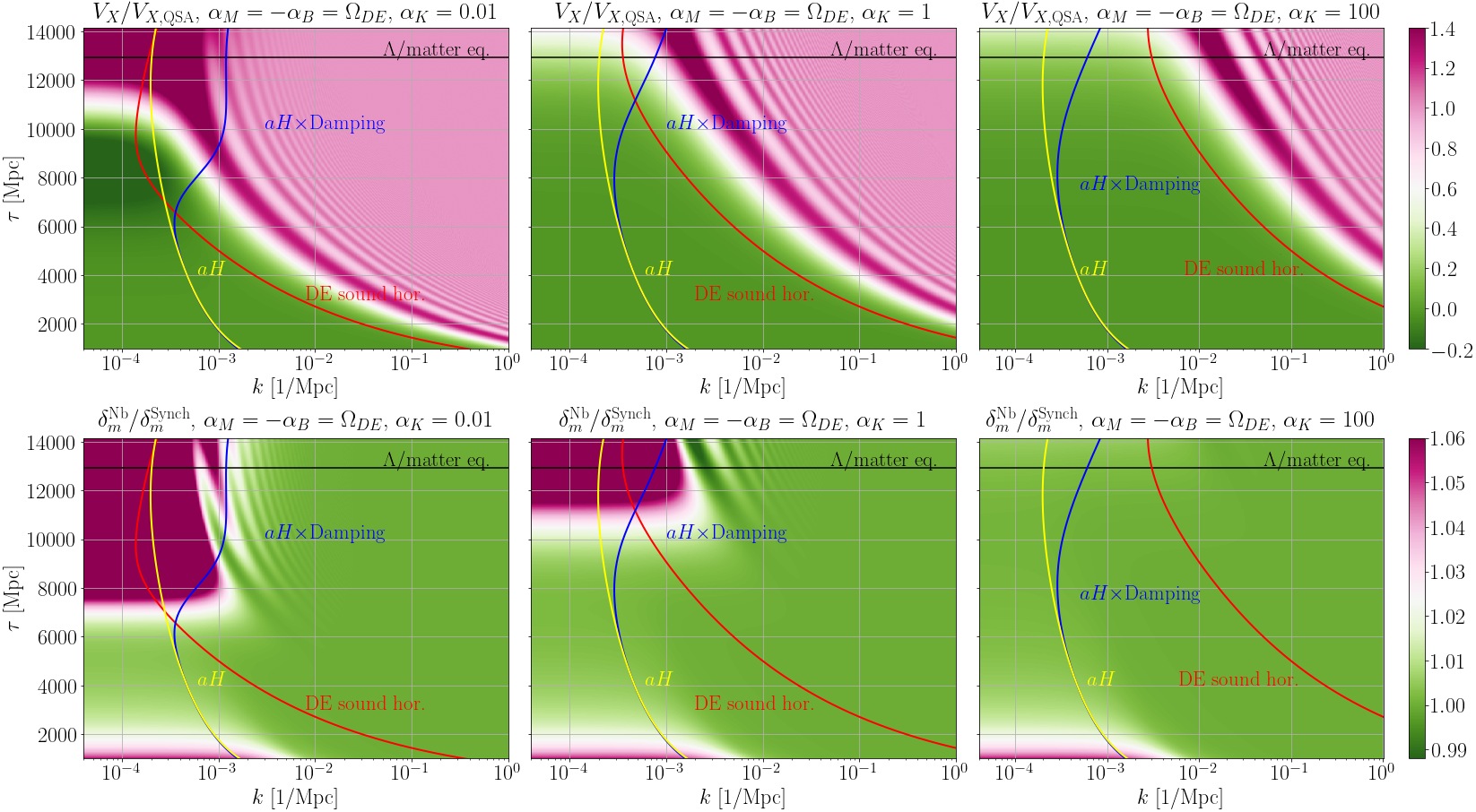} 
\caption{Scale and time dependence of GAOs. \textbf{Top row:} Two-dimensional $(k, \tau)$ plot of the ratio $V_{X}/V_{X}^{\rm QSA}$ in Jordan-Brans-Dicke gravity, for three different values of kineticity, $c_{\rm K} = 0.01, 1,100$. \textbf{Bottom row:} Two-dimensional plot of the ratio $\delta^{\rm Nb}_{\rm m}/\delta^{\rm S}_{\rm m}$ in the same theory, for three different values of kineticity, $\kin = 0.001, 1,100$. The dark energy sound horizon (red curve) Equation (\ref{eq:k_H}), the damping term (blue curve) in Equation (\ref{eq:metric_vx}) and the cosmological horizon (yellow curve) $aH$ are also plotted. The oscillations in the top row plots occur once a given $k$ mode crosses the dark energy sound horizon and damping scale, and, once inside this region, get slowly damped by the QSA contribution. These oscillations are seen in the matter density contrast for cases in which the crossing happens at scales much larger than the QSA regime at large $k$, as seen in the bottom left and center plots.}
\label{fig:k_tau_plot}
\end{figure}

To understand the origin of these oscillations, we also plot the evolution of the ratios $V_{X}/V_{X}^{\mathrm{QSA}}$ and $\delta_{\rm m}^{\mathrm{Nb}}/\delta_{\rm m}^{\mathrm{S}}$ as a function of wavenumber, $k$, and conformal time, $\tau$, in Figure~\ref{fig:k_tau_plot}. $V_{X}/V_{X}^{\mathrm{QSA}}$ minimises the non-oscillatory contributions from the QSA, while $\delta_{\rm m}^{\mathrm{Nb}}/\delta_{\rm m}^{\mathrm{S}}$ also allows us to highlight these features in the matter density contrast in the N-body gauge, since at late times and inside the horizon, the N-body gauge and the synchronous gauge are approximately the same. This highlights that any difference in behavior between the two is a purely relativistic effect.

From the scalar field fluctuation equation, (\ref{eq:metric_vx}), the only way acoustic waves may appear is when $V_{X}$ crosses the dark energy (DE) sound horizon, defined as:
\begin{equation}\label{eq:k_H}
      k_{\rm H} = \frac{a H \sqrt{2-\bra}}{c_{\rm s}\sqrt{2}}.
\end{equation}
When a given $k$ mode enters the sound horizon, pressure gradients from the scalar field act to counter balance the gravitational attraction. Therefore, as we can see in Figure~\ref{fig:k_tau_plot}, when the scalar field fluctuation of a specific Fourier mode crosses the sound horizon, gravity acoustic oscillations (GAOs) emerge.
%%%%%%%%%%-----------_%%%%%%%%%%%%%
%%%%%%%%%%-----------_%%%%%%%%%%%%%
These GAOs, however, are damped by two effects: the damping term multiplying $V_{X}^{\prime}$ in Equation (\ref{eq:metric_vx}), and when matter density perturbations start to dominate $V_{X}$. The former is represented by the blue lines in Figure~\ref{fig:k_tau_plot}, and we can see that modes must also be inside this scale to oscillate. 

\begin{figure}[h] 
\centering
\includegraphics[width=1.\textwidth]{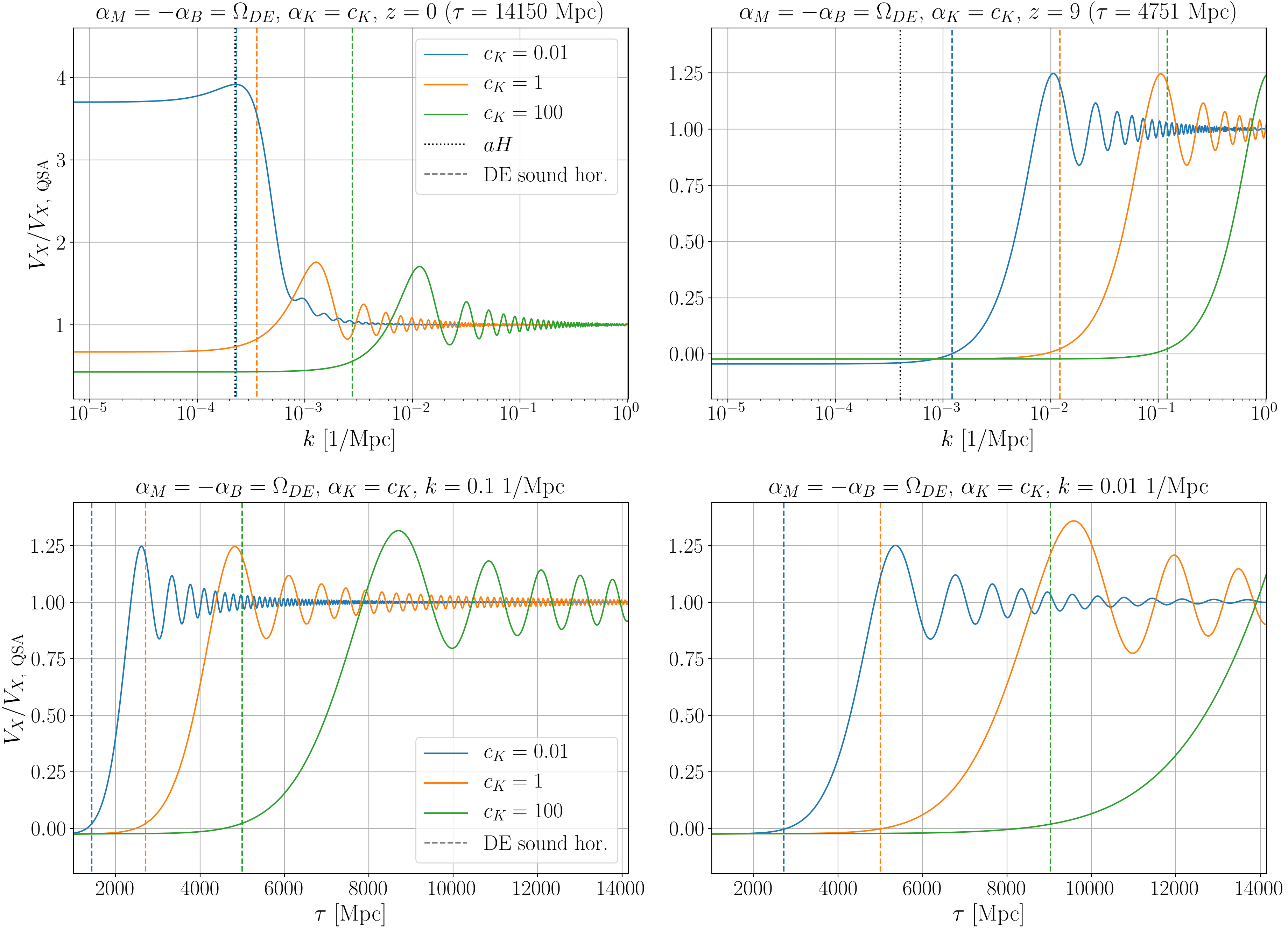} 
\caption{Scale and time dependence of GAOs. \textbf{Top row:} Ratio $V_{X}/V_{X}^{\rm QSA}$ as a function of scale at two different redshifts in the JBD model at $z=0$ and $z=9$. \textbf{Bottom row:} The same ratio, but as a function of conformal time for two fixed Fourie modes, $k=0.1$ Mpc$^{-1}$ (left) and $k=0.01$ Mpc$^{-1}$ (right). We can see on the top row plots that the scalar field fluctuations evolve to the their QSA contribution once inside the sound horizon, dashed vertical lines, and their amplitude decreases as we go to smaller scales. In the bottom plots, the perturbation once it crosses the sound horizon grows and starts oscillating around its QSA value, getting damped with time, as the matter density perturbation contribution starts to dominate the full scalar field fluctuation. The sound horizon crossing happens at different wavenumbers and time in each model, as the sound horizon is affected by the value of the kineticity, seen from Equation (\ref{eq:k_H}).}
\label{fig:osc_Vx_over_Vx_QSA_of_tau_k}
\end{figure}

In Figure~\ref{fig:osc_Vx_over_Vx_QSA_of_tau_k} we illustrate the dependence of $V_{X}/V_{X}^{\mathrm{QSA}}$ on $k$ (at fixed $\tau$) and on $\tau$ (for fixed $k$) for the models shown in Figure~\ref{fig:k_tau_plot}. The upper plots show $V_{X}/V_{X}^{\mathrm{QSA}}$ as a function of scale at two different redshifts, $z=0$ and $z=9$, while the bottom plots present the same quantity as a function of conformal time for two specific Fourier modes, $k=0.1$ Mpc$^{-1}$ and $k=0.01$ Mpc$^{-1}$. The dashed coloured vertical lines represent the specific scale and conformal time of the dark energy sound horizon crossing for each model. And the black dotted vertical line is the conformal Hubble rate (the Hubble horizon). At large scales in the upper plots the full scalar field perturbation differs considerably from its QSA counterpart. And the same is seen in the bottom plots, where at early times $V_{X}$ is completely dominated by relativistic contributions. In all the plots we can see that once the perturbations cross the sound horizon they start oscillating about the QSA value.

Gravity acoustic oscillations are an intermediate-time effect, originating during matter domination. They are caused by the rapid evolution of the dark energy sound horizon, which at early times may be orders of magnitude smaller than the Hubble horizon, $(aH)^{-1}$. As we have seen in Figures~\ref{fig:JBD_plot} and \ref{fig:cm_plot}, the sound speed of the scalar field can start very small at early times, and then goes to order one values at late times, driving the evolution of $k_{\rm H}$. This makes modes that were outside the dark energy sound horizon cross inside the horizon, introducing pressure gradients and hence oscillations in the gravity sector.

From our previous discussion, $c_{\rm s}^{2}$ depends on the choice of parametrization of the kineticity function. In the present work we fixed this to be constant throughout the expansion of the Universe. However, we know from other works in the literature that gravity acoustic waves were not present if a different parametrization for $\kin$ was chosen. Specifically, if $\kin$ was proportional to the fractional dark energy density, $\Omega_{\mathrm{DE}}$, a common choice in the literature, we know that the sound speed of the scalar field is always of the same order in time, apart from a very brief interval at early times. Therefore, the dark energy sound horizon will exhibit a similar behavior to the cosmological Hubble horizon for most of the expansion history. While this discussion revolves around the use of parametrizations of Horndeski theories, in principle, we can find a covariant theory in which the sound speed evolves by orders of magnitude, specifically during matter domination, via an appropriate choice of the Horndeski functions $G_{i}$'s.

It is important to stress that GAOs do not affect the BAO peak in the matter power spectrum. This is due to the fact that the BAO scale lies inside the regime where the QSA already holds.

%%%%%%%%%%-----------_%%%%%%%%%%%%%
%%%%%%%%%%-----------_%%%%%%%%%%%%%
\section{Discussion}\label{sec:concl}
In this work we have presented the general implementation of the N-body gauge in Horndeski gravity. Following our previous investigation~\cite{us}, we have generalized the effective fluid description of modified gravity, in order to compute the relativistic density perturbation, $\delta \rho_{\mathrm{GR}}$, which can be included in modified Newtonian N-body simulations to make them consistent with relativistic perturbation theory on linear scales. We have implemented a numerical routine that uses the fluid equations of motion for non-pressureless matter and Equations (\ref{eq:drhoDE}-\ref{eq:sigmaDE}) to evaluate the terms in Equation (\ref{eq:k2gammanb}), in the public Einstein-Boltzmann code \hiclass \footnote{This implementation will be made available upon acceptance of this paper.}.

In Section~\ref{sec:sec2} we introduced the theoretical framework of our approach, with a brief review of the N-body gauge formalism and Horndeski's theory. The following section, Section~\ref{sec:MG_gen}, was devoted to the presentation of our main results. We showed the behavior of the relativistic corrections coming from modified gravity at large scales in four different modified gravity models, characterized by the $\alpha_{i}$ functions. The major conclusion from this investigation is the important role played by the kineticity function, $\kin$, enhancing, or suppressing, the signal at small wavenumbers; the smaller $\kin$, the bigger the signal. In Section~\ref{sec:MG_QSA} we showed how our formalism can be introduced in Newtonian N-body simulations of modified gravity. By separating the effects coming from the effective gravitational constant, $G_{\rm eff}$, using the QSA limit, and the ones coming from purely relativistic corrections, we showed that there are contributions to the matter power spectrum in modified gravity that are not captured by the usual N-body codes in such theories. As shown in Figures~\ref{fig:JBD_plot} and \ref{fig:cm_plot}, these effects can lead to effects greater than  $1\%$ in the matter power spectrum at scales where DESI and Euclid are expected to deliver below percent constraints. In some modified gravity models, further modifications to Newtonian N-body simulations are required due to the presence of screening mechanisms that suppress the modification of gravity on small scales. Since the screening mechanism operates on small scales and it does not affect large scale relativistic perturbations, this can be safely implemented in modified Newtonian simulations and our formalism can be used to make these simulations consistent with relativistic perturbation theory on linear scales. This argument assumes that the screening mechanism is effective on scales where the QSA approximation is valid. This is a reasonable assumption in all screening mechanisms studied in the literature.

The combination of Einstein-Boltzmann solvers with Newtonian N-body simulation codes is a fast and computationally low-cost method, and the introduction of the effective density perturbations, $\delta \rho_{\mathrm{GR}}$, in N-body simulations will allow us to interpret the output of simulations in a relativistic space-time. Consequently, one can perform ray-tracing techniques to construct the observed light-cone from simulations in a consistent manner.

In Section~\ref{sec:MG_osc} we discussed the presence of Gravity Acoustic Oscillations (GAOs) in the matter power spectrum and in the lensing potential, as shown in Figure~\ref{fig:mpk_osc}. These GAOs are caused by the dynamical nature of the additional scalar degree of freedom. In the models we considered, the GAO become significant due to the rapid evolution of the dark energy sound horizon, Equation (\ref{eq:k_H}), which is determined by the evolution of the scalar field sound speed, Equation (\ref{eq:cs2}). In the models we presented, $c_{\rm s}^{2}$ evolves from small values at early times to order one values at late times. The sound horizon of the scalar field is smaller than the cosmological horizon at early times, $k_{\rm H} \gg aH$, but at later times it becomes of the same order, as shown in Figure~\ref{fig:k_tau_plot}. The large variation of the dark energy sound horizon makes modes that were previously outside the horizon suddenly cross inside, which introduces pressure gradients that counter-act gravity. Once inside the horizon, a particular mode will oscillate until it is damped by the damping term in the scalar field fluctuation equation, (\ref{eq:metric_vx}), and by the contributions coming from matter density perturbations, which dominate as we move to greater values of $k$. This is shown in detail in Figure~\ref{fig:osc_Vx_over_Vx_QSA_of_tau_k}. Future LSS and 21cm surveys will be capable of probing scales in which the GAOs are observed, roughly $10^{-3}-10^{-2}$ Mpc$^{-1}$, and if we can detect the presence of such oscillations it could be a smoking gun for modified gravity.

Our results point towards a new possibility to constrain the kineticity function using future large-scale structure and 21cm intensity mapping surveys. However, the uncertainties associated with data coming from very large scales are still significant, due to cosmic variance. Multi-tracer techniques~\cite{uros,mcd_uros,abramo} can help us increase the constraining power coming and, therefore, a consistent study on how to combine our formalism with these methods is left for future works.

\acknowledgments
GB acknowledges support from the State Scientific and Innovation Funding Agency of Esp\'irito Santo (FAPES, Brazil) and the Coordena\c{c}\~ao de Aperfei\c{c}oamento de Pessoal de N\'ivel Superior - Brasil (CAPES) - Finance Code 001. KK and DW are supported by the UK STFC grant ST/S000550/1. KK is also supported by the European Research Council under the European Union's Horizon 2020 programme (grant agreement No.646702 ``CosTesGrav"). IS was supported by European Structural and Investment Fund and the Czech Ministry of Education, Youth and Sports (Project CoGraDS - CZ.02.1.01/0.0/0.0/15\_003/0000437). EB acknowledges support from the European Research CouncilGrant No: 693024 and the Beecroft Trust. 

%%%%%%%%%%%%%%
\appendix 

\section{$\alpha$ and $\lambda$ functions}\label{sec:AppA}
We present here the definitions of the $\alpha_{i}$ ($i= \textrm{B, M, K, T}$) and the $\lambda_{i}$ functions , ($i=1,...,8$), shown in Section~\ref{sec:Horn}.
\begin{align}
M_{*}^{2}\equiv & 2\left(G_{4}-2XG_{4X}-\frac{H\phi^{\prime}XG_{5X}}{a}+XG_{5\phi}\right)\\
\alpha_{\textrm{M}}\equiv & \frac{\dd \ln M_{*}^{2}}{\dd\ln a}\\
H^{2}M_{*}^{2}\alpha_{\textrm{K}}\equiv & 2X\left(G_{2X}+2XG_{2XX}-2G_{3\phi}-2XG_{3\phi X}\right)\\
 & +\frac{12H\phi^{\prime}X}{a}\left(G_{3X}+XG_{3XX}-3G_{4\phi X}-2XG_{4\phi XX}\right)\nonumber \\
 & +12H^{2}X\left[G_{4X}-G_{5\phi}+X\left(8G_{4XX}-5G_{5\phi X}\right)+2X^{2}\left(2G_{4XXX}-G_{5\phi XX}\right)\right]\nonumber \\
 & +\frac{4H^{3}\phi^{\prime}X}{a}\left(3G_{5X}+7XG_{5XX}+2X^{2}G_{5XXX}\right)\nonumber \\
HM_{*}^{2}\alpha_{\textrm{B}}\equiv & \frac{2\phi^{\prime}}{a}\left(XG_{3X}-G_{4\phi}-2XG_{4\phi X}\right)+8HX\left(G_{4X}+2XG_{4XX}-G_{5\phi}-XG_{5\phi X}\right) \label{eq:aB}\\
 & +\frac{2H^{2}\phi^{\prime}X}{a}\left(3G_{5X}+2XG_{5XX}\right)\nonumber \\
M_{*}^{2}\alpha_{\textrm{T}}\equiv & 4X\left(G_{4X}-G_{5\phi}\right)-\frac{2}{a^{2}}\left(\phi^{\prime\prime}-2aH\phi^{\prime}\right)XG_{5X}\,.
\end{align}
Each of these functions is independent of the others, and each has different physical meanings.

The $\lambda_{i}$ functions are:

\begin{align}
D= & \alpha_{\textrm{K}}+\frac{3}{2}\alpha_{\textrm{B}}^{2}\\
\lambda_{1}= & \alpha_{\textrm{K}}\left(1+\alpha_{\textrm{T}}\right)-3\alpha_{\textrm{B}}\left(\alpha_{\textrm{M}}-\alpha_{\textrm{T}}\right)\\
\lambda_{2}= & -\frac{3\left(\rho_{\textrm{m}}+p_{\textrm{m}}\right)}{H^{2}M_{*}^{2}}-\left(2-\alpha_{\textrm{B}}\right)\frac{H^{\prime}}{aH^{2}}+\frac{\alpha_{\textrm{B}}^{\prime}}{aH}\\
\lambda_{3}= & -\frac{1}{2}\left(2+\alpha_{\textrm{M}}\right)D-\frac{3}{4}\alpha_{\textrm{B}}\lambda_{2}\\
\lambda_{4}= & \alpha_{\textrm{K}}\lambda_{2}-\frac{2\alpha_{\textrm{K}}\alpha_{\textrm{B}}^{\prime}-\alpha_{\textrm{B}}\alpha_{\textrm{K}}^{\prime}}{aH}\\
\lambda_{5}= & \frac{3}{2}\alpha_{\textrm{B}}^{2}\left(1+\alpha_{\textrm{T}}\right)+\left(D+3\alpha_{\textrm{B}}\right)\left(\alpha_{\textrm{M}}-\alpha_{\textrm{T}}\right)+\frac{3}{2}\alpha_{\textrm{B}}\lambda_{2}\\
\lambda_{6}= & \left(1-\frac{3\alpha_{\textrm{B}}H^{\prime}}{\alpha_{\textrm{K}}aH^{2}}\right)\frac{\alpha_{\textrm{K}}\lambda_{2}}{2}-\frac{DH^{\prime}}{aH^{2}}\left[2+\alpha_{\textrm{M}}+\frac{H^{\prime\prime}}{aHH^{\prime}}\right]-\frac{2\alpha_{\textrm{K}}\alpha_{\textrm{B}}^{\prime}-\alpha_{\textrm{B}}\alpha_{\textrm{K}}^{\prime}}{2aH}-\frac{3\alpha_{\textrm{K}}p_{\textrm{m}}^{\prime}}{2aH^{3}M_{*}^{2}}\\
\lambda_{7}= & \frac{D}{8}\left(2-\alpha_{\textrm{B}}\right)\left[4+\alpha_{\textrm{M}}+\frac{2H^{\prime}}{aH^{2}}+\frac{D^{\prime}}{aHD}\right]+\frac{D}{8}\lambda_{2}\\
\lambda_{8}= & -\frac{\lambda_{2}}{8}\left(D-3\lambda_{2}+\frac{3\alpha_{\textrm{B}}^{\prime}}{aH}\right)+\frac{1}{8}\left(2-\alpha_{\textrm{B}}\right)\left[\left(3\lambda_{2}-D\right)\frac{H^{\prime}}{aH^{2}}-\frac{9\alpha_{\textrm{B}}p_{\textrm{m}}^{\prime}}{2aH^{3}M_{*}^{2}}\right]\label{eq:lambda_8}\\
 & -\frac{D}{8}\left(2-\alpha_{\textrm{B}}\right)\left[4+\alpha_{\textrm{M}}+\frac{2H^{\prime}}{aH^{2}}+\frac{D^{\prime}}{aHD}\right]\nonumber \\
c_{\text{sN}}^{2}= & \lambda_{2}+\frac{1}{2}\left(2-\alpha_{\textrm{B}}\right)\left[\alpha_{\textrm{B}}\left(1+\alpha_{\textrm{T}}\right)+2\left(\alpha_{\textrm{M}}-\alpha_{\textrm{T}}\right)\right]\,,
\end{align}
where $c_{\text{sN}}^{2}$ is the numerator of the sound speed squared of the
scalar field
\begin{equation}
\label{def:csN}
c_{\text{s}}^{2}=\frac{c_{\text{sN}}^{2}}{D}\,.
\end{equation}

%%%%%%%%%%%%%%%%%%%%%%%%%%%%%%%%% 

\end{document}